\documentclass{JHEP3}

\usepackage{epsfig}
\usepackage{graphicx}
\usepackage{enumerate}

\title{States and Boundary Terms: Subtleties of Lorentzian AdS/CFT}

\author{Donald Marolf\\

Physics Department, UCSB, Santa Barbara, CA 93106.

\texttt{marolf@physics.ucsb.edu}}

\abstract{We complete the project  of specifying the Lorentzian
AdS/CFT correspondence and its approximation by bulk
semi-classical methods begun by earlier authors.  At the end, the
Lorentzian treatment is self-contained and requires no analytic
continuation from the Euclidean.  The new features involve a
careful study of boundary terms associated with an initial time
$t_-$ and  a final time $t_+$.  These boundary terms are
determined by a choice of quantum states. The main results in the
semi-classical approximation are 1) The times $t_\pm$ may be
finite, and need only label Cauchy surfaces respectively to the
past and future of the points at which one wishes to obtain CFT
correlators.  Subject to this condition on $t_\pm$, we provide a
bulk computation of CFT correlators that is manifestly independent
of $t_\pm$.   2) As a result of (1), all CFT correlators can be
expressed in terms of a path integral over regions of spacetime
{\it outside} of any black hole horizons. 3) The details of the
boundary terms at $t_\pm$ serve to guarrantee that, at  leading
order in this approximation, any CFT one-point function is given
by a simple boundary value of the classical bulk solution at null
infinity, $I$.   This work is dedicated to the memory of Bryce S.
DeWitt.  The remarks in this paper largely study the relation of
the AdS/CFT dictionary to the Schwinger variational principle,
which the author first learned from DeWitt as a Ph.D. student.  }

\date{November, 2004}


\begin{document}

\section{Introduction}

The discovery \cite{Juan} of gauge/gravity dualities has had a
dramatic effect on mathematical physics, and has led to new
approaches to understanding such diverse issues as QCD confinement
and the QCD spectrum, black hole entropy ,  and the microscopic
structure of quantum spacetime.  Some reviews can be found in
\cite{MAGOO,KlebRev,JuanRev}. One is also hopeful that it will
shed light on intrinsically dynamical issues such as black hole
evaporation and the possible formation and resolution of
singularities.

From the beginning \cite{Juan,GKP} there has been interest in
exploring the correspondence in Lorentzian signature.  However,
the Euclidean setting is somewhat simpler due to the lack of
propagating states. This simplicity was used in \cite{Witten} and
later works to flesh out various details of the correspondence.
Nevertheless, the importance of understanding Lorentzian AdS/CFT
remained clear, and a program to better formulate the Lorentzian
correspondence was pursued by Balasubramanian, Lawrence, and Kraus
\cite{BLK} and continued in work with Trivedi \cite{BLKT}.
Extensions to the case of black holes in equilibrium were also
studied \cite{KV,DKVK,SS,CS,KOS,LS}. Since the limit of large 't Hooft
coupling in the CFT corresponds to the classical limit in the
bulk, these works also pursued the important goal of understanding
how semi-classical methods in the Lorentzian AdS bulk can be used
to approximate CFT correlation functions. Indeed, this
approximation scheme has been the main avenue through which the
correspondence has been explored to date.

Of course, the interesting feature of the Lorentzian context (and
a central focus of some studies, e.g., \cite{BL,BR}) is the
existence of non-trivial propagating states.  As recognized in
\cite{BLK}, the corresponding wavefunctions must be properly
inserted into any path integral and will thereby determine the
correlation functions.  However, \cite{BLK,BLKT} did not study
this procedure in detail, and in particular the dependence (or
independence) of such wavefunctions on the boundary conditions at
null infinity (see Fig. \ref{AdSfig}) was not specified.  This
clearly leads to ambiguities when one wishes to compute CFT
correlators by varying the bulk boundary conditions\footnote{Such
ambiguities were recognized in \cite{BLKT} and were discussed
briefly in Appendix A of that work.}.
\FIGURE{\epsfig{file=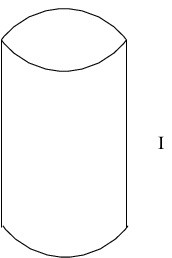, width=1in}\caption{We remind the
reader that null infinity $I$ consists of the endpoints of all
null geodesics.  In the AdS case, it is precisely familiar
cylindrical boundary indicated in the figure.} \label{AdSfig}}

Below, we show that such wavefunctions must be considered to be
independent, in either an advanced or retarded sense,  of the
boundary conditions at null infinity.  More specifically, we show
that to compute matrix elements of CFT operators between two
states, one must  vary the bulk boundary conditions at null
infinity while holding fixed the bulk wavefunction of one state in
an advanced sense and simultaneously fixing the bulk wavefunction
of the other state in a retarded sense.  Thus, we will see that
the Lorentzian AdS/CFT correspondence takes precisely the form of
the so-called Schwinger variational principle
\cite{Schwing,Bryce}.  This resolves the above ambiguity and
allows us to work out the detailed rules for the associated bulk
semi-classical approximation.  In contrast to \cite{SS,CS}, our
goal is precisely to formulate the Lorentzian correspondence in
the most general semi-classical context.  We find subtle
differences from the detailed prescription suggested in
\cite{BLKT} which can lead to significant changes (outlined below)
in situations associated with propagating bulk states.   An
example of such a case is the recent AdS/CFT analysis \cite{BL} of
inner horizon instabilities by Balasubramanian and Levi.

As in \cite{BLK,BLKT}, we proceed below by analytically continuing
the correspondence from the Euclidean setting, where the above
concerns do not arise.  Of course, one difference from
\cite{BLK,BLKT} will be a careful analysis of the manner in which
propagating states enter the story.  To remind the reader of
various subtleties and to fix notation, we first review the
analytic continuation of a standard local quantum field theory in
section \ref{ac}.    We then apply parallel arguments to the
AdS/CFT dictionary in section \ref{map}, arriving at the
conclusion stated above that the wavefunction of one state is to
be fixed in an advanced sense while the other is held fixed in a
retarded sense.  Though we argue by analytic continuation from the
Euclidean, we emphasize that the final result is a self-contained
Lorentzian prescription. The main differences from the
prescriptions of \cite{GKP,BLKT} are
\begin{enumerate}

\item CFT correlators associated with boundary points $x_1,\ldots,
x_n$ may be computed via a path integral over any region of
spacetime bounded by bulk  surfaces $\Sigma_\pm$ such that
$\Sigma_+$ ($\Sigma_-$) is a Cauchy surface for the bulk region to
the future (past)  of all points $x_i$.  In other words, the path
integral need only refer to the wedge regions described in
\cite{R1,R2}.

\item  As a result of (1), all CFT correlators can be expressed in terms of a path integral over regions of spacetime {\it outside} of any black hole horizons.
\end{enumerate}
At each stage our arguments are formal, though they mirror more rigorous arguments which may be applied when one considers non-gravitating quantum field theories on backgrounds of sufficient symmetry.  Even in the context of string theory, such arguments will hold in any approximation based on expanding about a classical solution.

The derivations of the above results are spelled out in sections
\ref{ac} and \ref{map}, but the conclusion can be quickly reached
by assembling the following observations: First, in the Euclidean
context it has previously been argued \cite{R2,R3} that quantum
fields in the CFT are essentially restrictions of quantum fields
in the bulk to the boundary of AdS (see also \cite{last}). Second,
this feature is naturally maintained under analytic continuation,
so that in the Lorentzian context the CFT operators are again the
restriction of the bulk operators to the boundary of AdS at null
infinity. Third, time-ordered correlators  $ \langle \beta | {\cal
O}_n(t_n) \ldots {\cal O}_1(t_1) | \alpha \rangle$ with $t_n \ge
\ldots \ge t_1$ in a Lorentzian field theory may be generated by
variations of the inner product $\langle \beta | \alpha \rangle$
which hold $ | \beta \rangle, |\alpha \rangle$ fixed respectively
at any finite future time $t_+ \ge t_n$ and at any past time $t_-
\le t_1$. This latter result is the Schwinger variational
principle \cite{Schwing,Bryce}.

Readers convinced by the short argument above, at least in
contexts where the bulk semi-classical approximation is valid, may
skip directly to section \ref{semi}.  Here the consequences for
the bulk semi-classical approximation are explored in more detail.
The main differences of the resulting semi-classical prescriptions
from that of \cite{GKP,BLKT} are
\begin{enumerate}
\item  Boundary terms at $t_\pm$ contribute in an essential way.
This is the case even for the vacuum state.

\item We see explicitly how the appropriate ``bulk/boundary propagator" is determined by the full quantum state.

\item  Results (1) and (2) interact in just such a way that,
 at  leading order in this approximation, any CFT one-point function is given by a simple boundary value of the classical bulk solution at null infinity, $I$.  This result holds regardless of the presence or absence of black hole horizons.
\end{enumerate}
Our prescription agrees with that of \cite{CS} in the particular context studied in that reference.

Section \ref{semi} derives these results in a general context and then explores them in a simple toy model which treats the bulk as a scalar field theory on a fixed AdS background.  Following this treatment,  section \ref{disc} summarizes the results and discussions their implications, especially for calculations along the lines of \cite{BL}.  We also discuss the extent to which our conclusions may be modified when quantum effects in the bulk gravitational field play an important role.

\section{Preliminaries: Analytic continuation in field theory}

\label{ac}

Here we review standard results on the analytic continuation of field theories between Lorentzian and Euclidean signatures.  Our arguments below are always formal, but mirror the basic structure of rigorous results (see, e.g., \cite{GJ}).  We begin in section \ref{LS1} with a review of the correspondence between correlation functions and variations of path integrals in standard Lorentzian quantum field theories.  This provides the opportunity to fix notation and to emphasize various subtleties which will be of particular use later in section \ref{map}.  We then review the analytic continuation to the Euclidean in section \ref{AC}.  Although such results are familiar in this context, stating them explicitly will allow us to carry them over directly to the AdS/CFT case of interest in section \ref{map}.

\subsection{Lorentzian Field Theory: Notation and the Schwinger variational principle}

\label{LS1}

We begin with a brief review of Lorentzian field theory which we use to establish our notation and conventions.
In particular, a pedagogical introduction to the Schwinger variational principle \cite{Schwing,Bryce} is provided.  This will be of use  in section \ref{map}.

Let us consider a quantum field theory with a Hilbert space of states and operators ${\cal O}(t)$ that are local in (Lorentzian) time.  To avoid clutter in our notation, the dependence of our operators on space will seldom be explicitly displayed.  Although analytic continuation is of most interest for theories with time-translation symmetry, it will be useful to momentarily allow our theory to have an arbitrary time-dependent Hamiltonian, given at time $t$ by a self-adjoint operator $H_t$.  We work in the Heisenberg picture, so the statement that our theory admits a time-dependent Hamiltonian is equivalent to the statements
\begin{eqnarray}
\label{tEv}
\frac{d}{dt} {\cal O}(t) &=& i [H_t, {\cal O}(t)] \ \ \ {\rm or} \cr
{\cal O}(t_2) &=& U(t_2,t_1) {\cal O}(t_1) U^{-1}(t_2,t_1)  \ \ \ {\rm where} \ \ \
U(t_2,t_1) = {\cal P} \exp\left( i \int_{t_1}^{t_2} H_t dt \right),
\end{eqnarray}
where ${\cal P}$ denotes path ordering with operators associated to the final end of the path appearing most to the left.  We have chosen to indicate the time dependence of $H_t$ by a subscript in order to emphasize the difference between the dependence of $H_t$ on $t$ (which is in principle arbitrary  as it may depend on parameters external to our system) and the $t$-dependence (\ref{tEv}) of the local operators ${\cal O}(t)$.  We shall endeavor to preserve the notation $A(t)$ for operators whose time-dependence is given by (\ref{tEv}) while denoting arbitrarily specifiable time-dependence by a subscript.  Note that we take $H_t$ to be Hermitian, so that ${\cal O}(t)$ is Hermitian if it is Hermitian at any time $t'$.

A case of particular interest occurs when $H_t$ takes the form
\begin{equation}
\label{sourceH}
H_t = H_{J_t}(t) := H^0(t) + \sum_i J^i_t {\cal O}_i(t),
\end{equation}
where the time-dependence of $H^0(t)$ and the ${\cal O}_i(t)$ are given by (\ref{tEv}) but the `sources' $J^i_t$ are arbitrarily specifiable classical functions of $t$.  In other words, we have
\begin{equation}
\label{SH}
H_t = U(t,0)[ H^0(0) + J^i_t {\cal O}_i(0)] U^{-1}(t,0),
\end{equation}
where we note  that $U(t,0)$ can be calculated from knowledge only of $H_{t'}$ for $t > t' > 0$.

In principle, one must solve (\ref{SH}) and (\ref{tEv}) simultaneously to construct $H_t$ and $U(t,0)$, though we will display a shortcut below.

In the particular case where $J=0$, (\ref{SH}) reduces to the time-independent case $H_t = H^0(t) = H^0(0)$.
We will often use the notation $J_t$ (with the $i$ index supressed) to indicate the collection of all sources at a given time $t$, and the notation $J$ (with the $t$ subscript suppressed) will indicate the collection of all functions $J^i_t$.   Below we will always take $H_t$ to be of the form (\ref{sourceH}), with $H^0(0)$ (and thus $H^0(t)$) a positive semi-definite operator.

Note that, once the sources $J^i_t$ are specified at the time $t$,  equation (\ref{sourceH}) provides a recipe for constructing the Hamiltonian $H_t$ from the local operators at time $t$.  The notation in (\ref{sourceH}) indicates that we have chosen to denote this recipe by $H_{J_t}(t)$.  This notation admits an obvious generalization, in that we may also apply the recipe given by $J_{t_1}$ to the local operators associated with some other time $t_2$:
\begin{equation}
\label{2t}
H_{J_{t_1}}(t_2) := H^0(t_2) + \sum_i J^i_{t_1} {\cal O}_i(t_2).
\end{equation}
Here again, the fact that we have used parentheses for the argument $t_2$ indicates that these operators satisfy
\begin{equation}
\label{3t}
H_{J_{t_1}}(t_2)   = U(t_2,t'_2)  H_{J_{t_1}}(t'_2) U^{-1}(t_2,t'_2) .
\end{equation}
In particular, equation (\ref{SH}) above is an example of (\ref{3t}) at $t'_2=t$ and $t_2=0$.

The two-time construction (\ref{2t}) will be surprisingly useful in our discussion below.  This occurs because the evolution operator $U(t_+,t_-)$ may be written in a useful way in terms of $H_{J_t}(t_-)$ and thus in terms of operators associated with a {\it single} time.  To see this, one need only rewrite the defining differential equation for $U(t,t_-)$ as
\begin{eqnarray}
\frac{dU(t,t_-)}{dt} &=& iH_{J_t}(t) U(t, t_-)=  iU(t, t_-)\Bigl( U^{-1}(t, t_-)H_{J_t}(t) U(t, t_-) \Bigr) \cr
&=& i U(t, t_-) H_{J_t}(t_-), \ \ \ {\rm so \ that ,} \cr
U(t,t_-) &=& {\cal P}^{-1}  \exp\left( i \int_{t_-}^{t_+} H_{J_t}(t_-) dt \right),
\end{eqnarray}
where ${\cal P}^{-1}$ denotes inverse path-ordering.  This provides a solution for $U(t,t_-)$ in terms only of operators at the single time $t_-$.

We will be interested in the vacuum states $|0; t\rangle_J $ of the operators $H^0(t)$, satisfying $H^0(t)|0; t\rangle_J =0$, and we assume that $H^0(t)$ is such that there is a unique such state (up to a phase).   The phase of such states is of course arbitrary, but it is convenient to choose the phases to satisfy
\begin{equation}
\label{tEv2}
|0;t \rangle_J = U(t,t') |0; t' \rangle_J.
\end{equation}
Here the subscript $J$ denotes the implicit dependence of the state on the sources through $U(t,t')$.
The inner product ${}_J\langle 0;t_+ | 0; t_- \rangle_{J}$ of two such vacuum states at times $t_\pm$ also depends on the sources $J$.

To express this dependence in a useful form, let us pick an arbitrary reference time $t=0$ and use the operators ${\cal O}(0)$ to define an isomorphism between the quantum theories defined by different sources $J$ for the same operators $H^0(0)$.  That is, we take operators ${\cal O}(0)$ and their eigenstates to be independent of $J$, while operators at all other times depend on $J$.  Thus we may write $|0; t=0 \rangle = |0; t=0 \rangle_J$, indicating that, in the sense defined by these isomorphisms, the vacuum of $H^0(0)$ is independent of the choice of source $J$.

The state $|0; t=0 \rangle$ may now be used to write the inner product between two states $|0; t_\pm \rangle$, which are vacuua of $H^0(t_\pm)$, in a form which manifestly displays  the dependence on the source $J$:
\begin{eqnarray}
\label{expJ}
{}_J \langle 0; t_+ | 0; t_- \rangle_J &=&  \langle 0; t=0| U^{-1}(t_+,0) U(t_-,0) | 0; t=0 \rangle \cr
&=&  \langle 0; t=0|
{\cal P}  \exp\left(- i \int_{0}^{t_+} H_{J_t}(0) dt \right)
{\cal P}^{-1}  \exp\left( i \int_{0}^{t_-} H_{J_t}(0) dt \right)
| 0; t=0 \rangle \cr
&=&
\langle 0; t=0|
{\cal P}  \exp\left(- i \int_{t_-}^{t_+} H_{J_t}(0) dt \right)
| 0; t=0 \rangle.
\end{eqnarray}
Note that the final form of (\ref{expJ}) depends on the sources $J$ only through
$H_{J_{t}}(0) := H^0(0) + \sum_i J^i_{t} {\cal O}_i(0),$ since all other quantities at $t=0$ have been {\it defined} to be independent of $J$.

Having clarified the dependence of this inner product on the sources $J$, we may vary $J(t)$ for $t_+ > t > t_-$ and use the above inner product as a generating functional for correlators.  Using the final form of (\ref{expJ}) and taking $t_+ > t_n > \ldots t_1 > t_-$ one finds
\begin{eqnarray}
\label{corr}
&&\left( i \frac{\delta}{\delta J^n_{t_n}} \right)
 \ldots  \left( i \frac{\delta}{\delta J^1_{t_1}}  \right)  {}_J \langle 0; t_+ | 0; t_- \rangle_J \cr
&&=
\langle 0; t=0|{\cal P}  e^{\left(- i \int_{t_n}^{t_+} H_{J_t}(0) dt \right)}
{\cal O}_n(0) {\cal P}  e^{\left(- i \int_{t_{n-1}}^{t_n} H_{J_t}(0) dt \right)}
\ldots {\cal O}_1(0) {\cal P}  e^{\left(- i \int_{t_{-}}^{t_1} H_{J_t}(0) dt \right)}
| 0; t=0 \rangle \cr
&&= {}_J\langle 0; t_+|
{\cal O}_n(t_n)
\ldots {\cal O}_1(t_1)
| 0; t_- \rangle_J,
\end{eqnarray}
where the last step follows from our time evolution equations (\ref{tEv}) and (\ref{tEv2}).

Thus, variations of the inner product ${}_J \langle 0; t_+ | 0; t_- \rangle_J$ with respect to the sources $J$ generate the associated time-ordered correlation functions.  If one wishes, one may use the standard skeletonization arguments (see e.g. \cite{HT,W}) to write this inner product as a path integral:
\begin{eqnarray}
\label{Lpath}
{}_J \langle 0; t_+ | 0; t_- \rangle_J &=& \int_{[t_-,t_+]} {\cal D} \phi e^{iS}
 \langle 0| \phi(t_+); 0;t=0  \rangle
\langle \phi(t_-); 0 | 0; t=0 \rangle,
 \cr
&=& \int_{[t_-,t_+]} {\cal D} \phi e^{iS}
 {}_J\langle 0; t_+| \phi(t_+); t_+  \rangle
\langle \phi(t_-); t_- | 0; t_- \rangle_J ,
\end{eqnarray}
where the action is given by appropriate interpretation of the expression:
\begin{equation}
\label{action} S = \int_{t \in [t_-,t_+]} (p \dot{q} - H_t) \ dt.
\end{equation}

Here $(p,q)$ are coordinates on the classical phase space associated with the fields $\phi$; e.g., with $q$ representing the configurations $\phi$ at each time $t$ and with the momenta $p$ representing certain functions of $\phi$ and the velocities $\dot \phi$.  The dot denotes a derivative with respect to $t$ and
and we understand that $H_t$ in (\ref{action}) is the classical phase space function obtained from the quantum operator $H_t$  by appropriately factor ordering configuration and momentum operators.  One could choose to interpret the above expressions as directly defining a phase space path integral (with ${\cal D}\phi$ in (\ref{Lpath}) replaced by ${\cal D}q {\cal D}p$), but we have in mind that  (\ref{Lpath}) will be a configuration space path integral so that the momenta $p$ should be replaced with their expressions in terms of $q,\dot{q}$ as determined by the corresponding equations of motion\footnote{We remind the reader that this last step is really just assuming that the momentum integrals have already been performed.  When the action is quadratic in momenta, this can be done exactly using the semi-classical approximation, though this may generate a non-trivial measure.  The more general case is more complicated, but can still be written in the above form by the appropriate choice of measure ${\cal D}\phi$.}.

The objects $|\phi; t \rangle$ are eigenstates of all (configuration) fields $\phi(t)$ at the time $t$ and have have eigenvalues $\phi$, where we take $|\phi; t \rangle$ to satisfy the analogue of equation (\ref{tEv2}).  Thus, the inner products in (\ref{Lpath}) represent wavefunctions of the states $|0;t=0 \rangle$ and $|0; t_\pm \rangle_J$ at times $t=0, t_\pm$.  Note that the wavefunctions in the first line are equal to those in the second line by the unitarity of our time evolution.  Which form of the path integral seems most natural depends on whether one prefers to perform the skeletonization in the Heisenberg picture (starting with ${}_J \langle 0; t_+ | 0; t_- \rangle_J $) or in the Schr\"odinger picture (starting with the expression $\langle 0; t=0|
{\cal P}  \exp\left(- i \int_{t_-}^{t_+} H_{J_t}(0) dt \right)
| 0; t=0 \rangle $).

The notation in (\ref{Lpath}) indicates that we integrate over all fields $\phi$ associated with the closed time interval $[t_-,t_+]$.  In particular, the integrations over $\phi(t_\pm)$ serve to attach the vacuum wavefunctions indicated above. Now, when considering field theory in Minkowski space, one can often ignore the details of these wavefunctions in the limit $t_\pm \rightarrow \pm \infty$ due to the fact that particles in Minkowski space disperse.  Thus, changing the wavefunctions tends to result in at most a change in normalization, which is usually not of interest.  However, more care will be required in the AdS/CFT context.  On a compact space or in AdS space, particles do not disperse and even after a long time any state can be distinguished from the vacuum.  As a result, the wavefunctions associated with the integrations over $\phi(t_\pm)$ will be important for our story.

Finally, it is clear from (\ref{corr}) and (\ref{Lpath}) that one may compute correlation functions of time-ordered products of operators by varying the path integral (\ref{Lpath}).  In this context, we point out a subtlety concerning the wavefunctions $\langle \phi(t_\pm);0|0;t=0 \rangle$ and
$\langle \phi(t_\pm);t_\pm|0 ;t_\pm \rangle_J$.  The wavefunction $\langle \phi(t_\pm);0|0;t=0 \rangle$ is independent of the sources $J$, so it is clear that it is held fixed under the variations if we use the first path integral expression in (\ref{Lpath}).    If on the other hand we use the second path integral expression in (\ref{Lpath}), then we must consider how the wavefunction $\langle \phi(t_\pm);t_\pm|0 ;t_\pm \rangle_J$ is to be varied with $J$.  But by the unitarity of our time evolution, these second wavefunctions are numerically equal (for all $J$) to the manifestly $J$-independent wavefunctions
$\langle \phi(t_\pm);0|0;t=0 \rangle$.  Thus, the wavefunctions at $t_\pm$ are similarly held fixed during variations of (\ref{Lpath}) which compute the correlators (\ref{corr}).

Note, however, that we may also consider wavefunctions of the states $ |0; t_\pm \rangle_J$ associated with some generic time $t$ by taking inner products with $|\phi; t\rangle$.  These inner products  $\langle \phi;t | 0 ;t_\pm \rangle_J$ {\it do} depend on the sources $J$ for $t < t_+$ or $t > t_-$ respectively.  Thus, we may say that variations of the vacuum to vacuum transition function
${}_J \langle 0; t_+ | 0; t_- \rangle_J$ are taken with $|0; t_+ \rangle$ being held fixed in an {\it advanced} sense (i.e., at time $t_+$), while $|0; t_- \rangle$ is held fixed in a {\it retarded} sense (i.e., at time $t_-$).  With this understanding (\ref{corr}) is just the ``Schwinger variational principle" \cite{Schwing,Bryce}, typically written in the form
\begin{equation}
\delta \langle \beta |  \alpha \rangle =  -i \langle \beta | \delta S|  \alpha \rangle,
\end{equation}
where $\delta S$ is regarded as an operator.  Our comments above are largely a pedagogical exposition of this principle in notation which will be convenient for the following sections.  This completes our review of the Lorentzian field theory and establishes our Lorentzian notation.

\subsection{Analytic Continuation}
\label{AC}

Recall that our goal is to carefully analytically continue the Euclidean AdS/CFT dictionary to Lorentz signature and to study the resulting consequences.  Since this prescription is typically given as a path integral, it is prudent to review the sense in which the more familiar path integral (\ref{Lpath}) can be analytically continued from Lorentzian to Euclidean signature.   Of course, the path integral (\ref{Lpath}) is not directly a function of time.  In reality, it is the correlation functions (\ref{corr}) whose arguments $t_n,\ldots, t_1$ one wishes to continue to imaginary values $t_n= -i \tau_n,\ldots, t_1= -i \tau_0$.  We remind the reader that the so-called analytically continued path integral is nothing other than a generating functional for the analytically continued correlation functions.

It is clear that a strict analytic continuation is possible only if the sources $J(t)$ are analytic functions.  In fact, it will be sufficient for our purposes to consider the case $J=0$.  Of course, one must allow $J$ to be non-zero during a variation, but we will simply be interested in the result of the variation evaluated at $J=0$.  In this case, we see from (\ref{corr}) that the correlation functions may be written in the form
\begin{eqnarray}
\label{ZLcorr}
&{}_{J=0}&\langle 0; t_+|
{\cal O}_n(t_n)
\ldots {\cal O}_1(t_1)
| 0; t_- \rangle_{J=0}, \cr
&=&
\langle 0; t=0|e^{-i ({t_+}-{t_n}) H^0(0)}
{\cal O}_n(0) e^{-i ({t_n}-{t_{n-1}}) H^0(0)}
\ldots {\cal O}_1(0) e^{-i ({t_1}-{t_-}) H^0(0)}
| 0; t=0 \rangle, \ \  \ \ \
\end{eqnarray}
where the analyticity in $t$ is now manifest in the domain  $Im(t_+ - t_n) \le 0$,  $Im(t_1 - t_-) \le 0$, $Im(t_i - t_j) \le 0$ for $i > j$.  We may thus
analytically continue this result to imaginary $t_i= - i \tau_i$:
\begin{eqnarray}
\label{Ecorr}
&{}_{J=0}&\langle 0: \tau_+|
{\cal O}_n[\tau_n]
\ldots {\cal O}_1[\tau_1]
| 0: \tau_- \rangle_{J=0},
\cr &=&
\langle 0; t=0|e^{- ({\tau_+}-{\tau_n}) H^0(0)}
{\cal O}_n(0) e^{- ({\tau_n}-{\tau_{n-1}}) H^0(0)}
\ldots {\cal O}_1(0) e^{- ({\tau_1}-{\tau_-}) H^0(0)}
| 0; t=0 \rangle, \ \ \ \ \
\end{eqnarray}
for $\tau_+ \ge \tau_n \ge \ldots \ge \tau_1 \ge \tau_-$.
Here the quantities on the left-hand side are by definition the analytic continuation of the corresponding quantities on the left-hand side of (\ref{ZLcorr}) with the understanding that ${\cal O}_i(0)$, $H^0(0)$, and $|0\rangle$ are (naturally) taken to be constants\footnote{This is so for the case where the ${\cal O}$ are scalar operators, which we assume below for simplicity.  The more general case differs only by the insertion of additional factors of $i$ associated with time components of tensors.}.  Note several subtleties of our notation:  In order to prevent confusion between Lorentzian time dependence and Euclidean time dependence, we have made the definitions
\begin{equation}
\label{continue}
|0 : \tau \rangle := | 0; t= - i \tau  \rangle,  \ \langle 0 : \tau | := \langle 0; t= - i \tau| = (|0 : -\tau \rangle)^\dagger,
\ {\rm and} \  {\cal O}[\tau] := {\cal O}(t= - i\tau),
\end{equation}
so that the Euclidean expressions differ from their Lorentzian counterparts in the choice of colon ($:$) versus semicolon ($;$) in the time-dependence of states and in the choice of square versus round brackets in the time-dependence of operators.  Finally, we note that the operators of the form $e^{-(\tau_i-\tau_j)H^0(0)}$ in (\ref{Ecorr}) are bounded since $\tau_i \ge \tau_j$ and $H^0(0)$ is positive semi-definite.

As in the Lorentzian context, it is useful to extend our formalism to consider sources $J^\tau$ which we take to be arbitrary real functions\footnote{Thus, we do not take the $J^\tau$ to be the analytic continuation of any interesting Lorentzian sources $J_t$.  Instead, we merely develop a parallel Euclidean formalism, with our real interest being the source-free case $J=0.$} of $\tau.$
We may do as follows, where the reader can see that this definition is consistent with (\ref{Ecorr}) in the special case $J=0$.  We take the Euclidean time evolution to be generated by a family of Hermitian operators $H^\tau$ through:
\begin{eqnarray}
\label{tauEv}
{\cal O}[\tau] = U[\tau, 0] {\cal O}(0) U^{-1}[\tau,0], &\ \ \  &
|0: \tau \rangle_J = U[\tau,0] |0 \rangle, \ \ \ {\rm and} \ \ \ {}_J\langle 0: \tau | = \langle 0| U^{-1}[\tau,0]\cr
 \ \ &{\rm where}& \ \ \
U[\tau_2,\tau_1] = {\cal P} \exp\left( - \int_{\tau_1}^{\tau_2} H^\tau d\tau \right).
\end{eqnarray}
Here one should recall that $U[\tau,0]$ is not unitary, and that if the ${\cal O}(0)$ are Hermitian then we have
$({\cal O}[\tau])^\dagger = {\cal O}[-\tau]$ while in general $\langle 0: \tau| = (|0: - \tau \rangle)^\dagger.$

We assume now that $H^\tau$ is of the form
\begin{equation}
H^\tau = H^{J^\tau}[\tau] := H^0[t] + \sum_i J_i^{\tau} {\cal O}_i[\tau].
\end{equation}
We will use the notation $H^{J^\tau}[0]$ in analogy with the Lorentzian case.
The same steps as in the Lorentzian context again show that the correlation functions (\ref{Ecorr}) for $\tau_+ > \tau_n >  \ldots  \tau_1 > \tau_-$ can be computed through variations with respect to the Euclidean sources $J^\tau$:
\begin{equation}
\label{EJcorr} \left( - \frac{\delta}{\delta J_n^{\tau_n}} \right)
 \ldots  \left( - \frac{\delta}{\delta J_1^{\tau_1}}  \right)  {}_J \langle 0: \tau_+ | 0: \tau_- \rangle_J
= {}_J\langle 0 : \tau_+|
{\cal O}_n[\tau_n]
\ldots {\cal O}_1[\tau_1]
| 0: t_- \rangle_J.
\end{equation}
In particular, for $J=0$, (\ref{EJcorr}) gives the analytic continuation of the $J=0$ Lorentzian correlation functions.  Despite the Euclidean signature, one sees just as in the Lorentzian case that the states ${}_J\langle 0:\tau_+|$ and $|0: \tau_- \rangle_J$ are held fixed in the sense of advanced and retarded boundary conditions respectively.

Also as in the Lorentzian case, one may skeletonize the inner product
$ {}_J \langle 0: \tau_+ | 0: \tau_- \rangle_J $ to obtain a path integral expression of the form
\begin{eqnarray}
\label{Epath}
 {}_J \langle 0: \tau_+ | 0: \tau_- \rangle_J  &=& \int_{[\tau_-,\tau_+]} {\cal D} \phi e^{-S_E}
 \langle 0| \phi[\tau_+]: 0:\tau=0  \rangle \
\langle \phi[t_-]: 0 | 0 :\tau=0\rangle_J ,
 \cr
&=& \int_{[\tau_-,\tau_+]} {\cal D} \phi e^{-S_E}
 {}_J\langle 0: \tau_+| \phi[t_+]; \tau_+  \rangle \
\langle \phi[\tau_-]: \tau_- | 0: \tau_- \rangle_J ,
\end{eqnarray}
where the Euclidean action is given by
\begin{equation}
\label{Eaction}
S_E = - \int_{\tau \in [\tau_-,\tau_+]} (ip \dot{q}
- H^\tau) \ d\tau
\end{equation}
and the notation is directly analogous to that used in the Lorentzian path integral (\ref{Lpath}).

One difference from the Lorentzian case is that one can generically dispense with the wavefunctions in (\ref{Epath}) by taking $\tau_\pm $ to $\pm \infty$, provided that one intends to set $J=0$ after taking a finite number of variations.  Because the limit $\lim_{\tau \rightarrow \infty} e^{-H^0(\tau_1) \tau}$ is just the projection operator $| 0: \tau_1 \rangle \langle 0: \tau_1 |$, the details of the boundary conditions at large $\tau$ tend to affect the answer only by an overall normalization factor.  As opposed to the analogous Lorentzian operation, this simplification is as useful in the contexts relevant to AdS/CFT as it is for field theories on Euclidean ${\cal R}^n$.
Thus, we may define the  partition function $Z_J$ to be given by any path integral of the form
\begin{equation}
\label{PF}
Z_J := \int_{(-\infty,\infty)} {\cal D} \phi e^{-S_E},
\end{equation}
and generate time-ordered correlation functions via
\begin{equation}
\label{Epart} {}_J\langle 0 : \tau_+| {\cal O}_n[\tau_n] \ldots
{\cal O}_1[\tau_1] | 0: t_- \rangle_J = Z_J^{-1} \left( -
\frac{\delta}{\delta J_n^{\tau_n}} \right)
 \ldots  \left( i \frac{\delta}{\delta J_1^{\tau_1}}  \right)  Z_J,
\end{equation}
for $\tau_+ \ge \tau_n \ge \ldots \tau_1 \ge \tau_-$.
This concludes our review of analytic continuation for standard field theories.

\section{The AdS/CFT dictionary}

\label{map}

The considerations of section \ref{ac} above are straightforward and familiar to most researchers.  But the corresponding AdS/CFT context, in which one wishes to compute CFT correlation functions using a bulk AdS prescription, is often considered to be more subtle.  The crucial difference is typically thought to be the fact that CFT correlation functions are computed via variations of an AdS bulk path integral with respect to {\it boundary conditions} instead of {\it sources} of the form (\ref{sourceH}), which are more familiar.  The main point of our argument below is that variations with respect to boundary conditions  in fact {\it define} operators and that, as a result, such boundary conditions may be treated in precisely the same manner as the more familiar sources.

This result is implicit in the original statement of the Schwinger
variational principle $\delta \langle \beta |  \alpha \rangle = -i
\langle \beta | \delta S|  \alpha \rangle$, in which $\delta S$ is
regarded as an operator.   Our reasoning in section \ref{EBCS}
verifies this in the context of Euclidean AdS/CFT via a formal
argument which we hope will be of pedagogical use.  Even in this
context the result is not new.  In fact,  various works
\cite{R2,R3} have shown that, for the boundary conditions of
interest to the usual AdS/CFT dictionary, $\delta S$ is  simply
related to the asymptotic form of a local field in the
bulk\footnote{These works proceed by matching the perturbation
series defined by varying boundary conditions with that defined by
sources coupled to these asymptotic values.} (see also \cite{last}
for earlier steps in this direction).

 After demonstrating the above result, we analytically continue to the Lorentzian setting in section \ref{AdSAC} to obtain vacuum correlators and then deduce the corresponding result for non-trivial states $|\alpha \rangle, |\beta \rangle.$  The result is simply that the Lorentzian AdS/CFT prescription is again an implementation of the Schwinger variational principle.  Readers comfortable with this result are recommended to proceed directly to section \ref{semi}, where the implications for the classical limit are explored.  There we will find subtle differences from the semi-classical prescription of  Ref. \cite{BLKT}.

\subsection{Euclidean Boundary Conditions are Sources}
\label{EBCS}

We begin with the prescription of \cite{Witten} for the Euclidean AdS/CFT correspondence, for which no issues of propagating states arise and for which the boundary conditions at $\tau_\pm$ can be determined using only the symmetries of the Euclidean AdS and CFT spacetimes.  The recipe of \cite{Witten} is simply
\begin{equation}
\label{Zs}
Z_J^{CFT} = Z_J^{AdS},
\end{equation}
with the prescription that a given source $J$ in the CFT corresponds to a given set of {\it boundary conditions} $J$ for the Euclidean functional integral defining the AdS partition function $Z_J^{AdS}$, but that the bulk dynamics (e.g., the equations of motion) for the AdS theory do not depend on $J$.  The details of precisely which boundary conditions are associated with a given source are determined by the physics of D3-branes as described in e.g., \cite{Witten,DF}.

We wish to state clearly that the manipulations below will treat the bulk partition function as if it were the partition function of some conventional field theory.  We will, in particular, introduce wavefunctions of bulk states $|\alpha\rangle$ defined by the overlaps with eigenstates $|\phi; t\rangle$ of local fields $\phi$.  The reader is correct to question this, as the existence of an AdS/CFT correspondence is expected to imply that the bulk AdS theory is {\it not} a local field theory in a conventional sense.  Some aspect of this issue may be associated with the fact that the Euclidean action for truncations of this theory to, e.g., supergravity are not bounded below.

Nevertheless, it appears that the bulk theory becomes a local field theory in the low energy limit.
Furthermore, there has been significant interest in using semi-classical calculations in this low energy theory to obtain approximate results for CFT correlators in the limit of large 't Hooft coupling.  As the purpose of this work is to clarify the prescription for computing such approximate results in Lorentizian AdS/CFT, we now pass to this low energy limit and so justify our treatment of the bulk as a local field theory below.

We would like to consider the partition function $Z_J^{AdS}$ as an object of the form
$ {}_J~\langle~0:~\tau_+| 0: \tau_- \rangle_J  $ as in (\ref{Epath}).  Our first task is to introduce the coordinate $\tau$ on Euclidean AdS space, which is not a priori given to us by the dictionary (\ref{Zs}).
This is straightforward to do, as we may take $\tau$ to be the parameter defined by any vector field which is asymptotically a hypersurface-orthogonal Killing field of spaces satisfying Euclidean AdS boundary conditions such that translations $\tau \rightarrow \tau + \Delta$  act freely.

To the extent that the bulk theory may be treated as a local field theory, the partition function $Z_J$ may be expressed as a path integral of the form (\ref{PF}) associated with some real-valued Euclidean action.  This action then defines a real ($\tau$-dependent) classical Hamiltonian via the usual Legendre transform implicit in (\ref{Eaction}).  At the quantum level, slicing the path integral along surfaces defined by $\tau_\pm$ defines a Euclidean time-evolution operator $U[\tau_+,\tau_-]$ though
\begin{eqnarray}
\label{cut}
{}_J \langle \phi_+: \tau_+ |  \phi_- : \tau_- \rangle_J  &:=& \int_{(\tau_-,\tau_+)} {\cal D} \phi e^{-S_E}, \ \ \ {\rm and} \cr
{}_J \langle \phi: \tau_- |  U^{-1}[\tau_+,\tau_-] |  \phi : \tau_- \rangle_J & := & {}_J \langle \phi_+: \tau_+ |  \phi_- : \tau_- \rangle_J,
\end{eqnarray}
where the notation in the first line indicates that we integrate only over fields $\phi[\tau]$ associated with $\tau$ in the {\it open} interval $(\tau_-,\tau_+)$.  We do not integrate over the boundary values $\phi(\tau_\pm)$ in (\ref{cut}); rather, the boundary values $\phi_\pm$ are specified by the states on the left-hand side whose overlaps we wish to compute.  This construction effectively also defines the Hilbert space associated with the quantum theory in terms of eigenstates $ |  \phi : \tau \rangle_J  $ of
operators $\phi[\tau]$ which satisfy
\begin{equation}
\phi [\tau] = U[\tau, 0] \phi (0) U^{-1}[\tau,0].
\end{equation}
The quantum Hamiltonian $H^\tau$  is then defined by the $\tau$-derivative of $U$ through
\begin{equation}
\label{defH}
H^\tau = - \left( \frac{\partial}{\partial \tau} U[\tau, \tau'] \right) U^{-1}[\tau, \tau'],
\end{equation}
where the result is independent of $\tau'$.  We shall assume that $H^\tau$ is Hermitian, in agreement with the corresponding operator in the CFT.

Note that because (\ref{defH}) is independent of the reference time $\tau'$, the action of  $H^\tau$
on the states $|\phi; \tau \rangle$ associated with the same value of $\tau$ can be computed knowing only the boundary conditions $J^\tau$ associated with this {\it same} value of $\tau$.  In this sense we may consider $H^\tau$ as a functional only of $J^{\tau'}$ for $\tau = \tau'$ and we may consider variations of $H^\tau$ with respect to $J^\tau$.  Note that the dictionary that relates AdS boundary conditions to CFT sources defines a preferred set of boundary conditions associated with $J=0$.   Thus, the variations about $J=0$ define operators
\begin{equation}
{\cal O}_i[\tau] := \frac{ \delta H^\tau}{\delta J_i^\tau} \big|_{J=0}.
\end{equation}
Here the label $i$ simply refers to a particular way in which the boundary conditions can be varied.  In essence, the index $i$ on $J_i^\tau$ is a vector index referring to the tangent space to the space of boundary conditions $J$ (and thus a dual-vector index\footnote{Our notation places this $i$ in the standard location in the Lorentzian theory, but we have found it convenient to write $i$ as a lower index $J_i^\tau$ in the Euclidean theory to produce a clean notation which distinguishes between Euclidean and Lorentzian quantities.} on ${\cal O}$).

We may expand the Hamiltonian $H^\tau$ in the form
\begin{equation}
H^\tau := H_{J^\tau}[\tau] = H^0[\tau] + \sum_i J_i^\tau {\cal O}_i[\tau] + \ {\rm terms \ of \ order} \ J^2,
\end{equation}
after which it is natural to refer to ${\cal O}_i[\tau]$ as the operator coupled to the {\it source} $J_i^\tau$.
 The only property of the $J=0$ boundary conditions which we will need is $\tau$-translation invariance, which is a symmetry of the CFT.  This guarantees that, for any $J$, we have
\begin{eqnarray}
H^0 [\tau] = U[\tau, 0] H^0[0] U^{-1}[\tau,0], \ \ \ {\rm and} \cr
{\cal O}_i[\tau]  = U[\tau, 0] {\cal O}_i[0]  U^{-1}[\tau,0]
\end{eqnarray}
as suggested by our notation.  Thus, we have all of the structure of section \ref{AC}.  In particular, we may also introduce the vacuum states $|0 : \tau \rangle_J$ of $H^0[\tau]$ satisfying $|0: \tau \rangle_J = U[\tau,0] | 0:0 \rangle_J$ and note that, in the case where $J$ vanishes sufficiently quickly as $\tau \rightarrow \pm \infty$, the partition function $Z_J$ effectively includes a factor of
\begin{equation}
\lim_{\tau \rightarrow \infty} e^{-H^0[\pm \infty]} = |0 : \tau = \pm \infty \rangle_J{}_J\langle 0 : \tau = \pm \infty |,
\end{equation}
which is the projection onto the associated vacuua\footnote{Here we assume that $H^0$ is positive semi-definite with a unique zero-energy eigenstate.  This is believed to be so for the case of relevance to the simplest AdS/CFT correspondence.  The more general case is a straightforward generalization.}.

Thus, when the sources $J^\tau$ vanish outside of some interval $(\tau_-,\tau_+)$, we may write the partition function as
\begin{equation}
\label{almost}
Z_J = N \ {}_J\langle 0 : \tau_+|0 : \tau_- \rangle_J,
\end{equation}
where $N$ is some undetermined normalization constant.   The important property of $N$ is that it does not depend on  $J^\tau$ for  $\tau \in (\tau_-,\tau_+)$.
Furthermore, we see that while the wavefunction ${}_J\langle 0 : \tau_+ | \phi : \tau \rangle$ for general $\tau$ depends on $J$, for $\tau \ge \tau_+$ this wavefunction is in fact independent of the boundary conditions $J^{\tau'}$ with $\tau' < \tau_+$, as the wavefunction was inserted by the projection operator defined by the functional integral over the region $\tau > \tau_+$.  Similarly, the wavefunction $\langle \phi: \tau |
0 : \tau_- \rangle_J$ is  independent of the boundary conditions $J^{\tau'}$ with $\tau' > \tau_-$.

In this sense, the vacuum wavefunctions satisfy the same advanced (retarded) boundary conditions as in (\ref{EJcorr}).  As a result, we may repeat all of the steps leading to (\ref{EJcorr}) to find as one might expect that normalized variations of the partition function yield correlation functions of $\tau$-ordered products of operators ${\cal O}_i[\tau]$:
\begin{eqnarray}
\label{EZJcorr}
\frac{1}{Z^{AdS}_J} \left( - \frac{\delta}{\delta J_n^{\tau_n}}\right)
 \ldots  \left( - \frac{\delta}{\delta J_1^{\tau_1}}  \right) Z^{AdS}_J &=&
 \cr
\left( - \frac{\delta}{\delta J_n^{\tau_n}} \right)
 \ldots  \left( - \frac{\delta}{\delta J_1^{\tau_1}}  \right)  {}_J \langle 0: \tau_+ | 0: \tau_- \rangle_J
&=& {}_J\langle 0 : \tau_+|
{\cal O}_n[\tau_n]
\ldots {\cal O}_1[\tau_1]
| 0: \tau_- \rangle_J,
\end{eqnarray}
for $\tau_+ \ge \tau_n \ge \ldots \ge \tau_1 \ge \tau_-$.
This provides a convenient way to rewrite the AdS/CFT dictionary (\ref{Zs}).  Introducing the operators $\tilde{\cal O}_i [\tau]$ to which the sources $J_i$ couple in the CFT and the associated vacuum states $|\tilde 0: \tau \rangle$ of the $J=0$ CFT Hamiltonian $\tilde H^0[\tau]$, we have
\begin{equation}
\label{mapEcors}
 {}_J\langle \tilde 0 : \tau_+|
\tilde {\cal O}_n[\tau_n]
\ldots \tilde  {\cal O}_1[\tau_1]
| \tilde 0: \tau_- \rangle_J
= {}_J\langle 0 : \tau_+|
{\cal O}_n[\tau_n]
\ldots {\cal O}_1[\tau_1]
| 0: \tau_- \rangle_J;
\end{equation}

i.e., the Euclidean AdS/CFT dictionary can be interpreted as mapping CFT correlators to appropriate bulk correlators.

\subsection{Analytic Continuation in AdS/CFT}
\label{AdSAC}

It is now straightforward to analytically continue this expression to Lorentz signature $t=-i\tau$.  As in section \ref{AC}, we most naturally think of (\ref{mapEcors}) as being analytically continued in the special case $J=0$.  However, since one simultaneously obtains expressions for all functional derivatives of the above correlators with respect to $J$, the more general expression
\begin{equation}
\label{mapLcors}
 {}_J\langle \tilde 0 ; t_+|
\tilde {\cal O}_n(t_n)
\ldots \tilde  {\cal O}_1(t_1)
| \tilde 0; t_- \rangle_J
= {}_J\langle 0 ; t_+|
 {\cal O}_n(t_n)
\ldots  {\cal O}_1(t_1)
|  0; t_- \rangle_J,
\end{equation}
holds at least at all orders in perturbation theory.
Now, if one desires, one may use the standard skeletonization arguments to express the right-hand size in terms of variations of a bulk AdS path integral:
\begin{eqnarray}
\label{mapLcors2}
 {}_J\langle \tilde 0 &;& t_+|\tilde {\cal O}_n(t_n)
\ldots \tilde  {\cal O}_1(t_1)
| \tilde 0; t_- \rangle_J  \cr \cr
&=& \left( i \frac{\delta}{\delta J^n_{t_n}} \right)
 \ldots  \left( i \frac{\delta}{\delta J^1_{t_1}}  \right)   \int_{[t_-,t_+]} {\cal D} \phi e^{iS^{AdS}} \
 {}_J\langle 0; t_+| \phi(t_+); t_+  \rangle
\langle \phi(t_-); t_- | 0; t_- \rangle_J , \ \
\end{eqnarray}
where again one holds $| 0; t_\pm \rangle_J$ constant in the sense of advanced (retarded) boundary conditions and we have assumed $t_+ \ge t_n \ge \ldots \ge t_1 \ge t_-$.

A key point is that, as stated thus far, the variations on the right are performed with respect to {\it sources} which couple to the operators ${\cal O}_i(t)$ defined by the procedure above, which we note originally defined these operators as the variation of a Euclidean path integral with respect to a set of boundary conditions.  Let us focus on the case $t=0$ for the simplest comparison with Euclidean expressions, since in fact ${\cal O}_i(0) = {\cal O}_i[0]$.  We see that the relevant boundary conditions are just those that define the operator $H_{J^0}[0] = H_{J^0}(0)$ and thus which define $H_{J_0}(0)$ .  As a result, performing the same variation of the corresponding Lorentzian path integral with respect to these boundary conditions leads to $-i$ times the variation of $H_{J_0}(0)$, which is just $-i {\cal O}_i(0)$.    Using time translations to make the analogous argument at any time, we find that the variations on the right-hand side of (\ref{mapLcors}) may be taken to be variations with respect to boundary conditions in {\it precisely}\footnote{Here we have treated all operators $\tilde {\cal O}_i$ as if they are scalars.  For the time components of tensor operators, additional factors of $i$ arise in the usual manner associated with analytic continuations.  The result is that the boundary conditions parametrized by $J$ in the Lorentzian bulk theory are just the natural analytic continuations of those in the Euclidean bulk theory.}
in the same manner as those implicit in the Euclidean AdS/CFT dictionary (\ref{Zs}).

Finally, it will be of interest to consider correlators  $ \langle
\tilde \beta |\tilde {\cal O}_n(t_n) \ldots \tilde  {\cal
O}_1(t_1) | \tilde \alpha \rangle$ involving non-trivial CFT
states $|\tilde \alpha \rangle, |\tilde \beta \rangle$ other than
the vacuum.  This is straightforward as by a clever choice of
sources $J_t$ for $t_+ > t > t_n$ and
 $t_1 > t > t_-$ we can in fact arrange for the wavefunctions $\langle \tilde \phi, t_\pm |\tilde 0 ; t_\pm \rangle_{J}$ to
match those of $\langle \tilde  \phi ; t_\pm |\tilde \alpha
\rangle $ and  $\langle \tilde  \phi ; t_\pm |\tilde \beta
\rangle$, so that these `vacuum' states match  our chosen states
exactly in the sense of retarded (advanced) boundary conditions
appropriate to (\ref{corr}).  Alternatively, one may think of
fixing $J$ and creating these states from the vacuum through the
action of appropriate operators $\tilde A(t),\tilde B(t)$
satisfying $|\tilde \alpha \rangle = \tilde A(t_-) | \tilde 0; t_-
\rangle_J$ and $|\tilde \beta \rangle = \tilde B(t_+) | \tilde  0;
t_+ \rangle_J$.  Such $\tilde A(t), \tilde B(t)$ are merely
examples of the general operators $\tilde {\cal O}_i(t)$ already
discussed, and so define operators $A(t),B(t)$ and states
$|\alpha\rangle= A(t_-) | 0; t_- \rangle_J$ and $|\beta \rangle=
A(t_-) | 0; t_- \rangle_J$ in the AdS bulk theory.  These
operators must correspond to the analogous variations of the bulk
path integral (\ref{mapLcors}) which, if taken at time $t_\pm$,
act directly on the vacuum states $|0;t_\pm \rangle$.  Thus, we
immediately have the result
\begin{eqnarray}
\label{abcors}
&& \langle \tilde \beta | \tilde {\cal O}_n(t_n)
\ldots \tilde  {\cal O}_1(t_1)
| \tilde \alpha \rangle  =
 \langle  \beta | {\cal O}_n(t_n)
\ldots  {\cal O}_1(t_1)
| \alpha \rangle \cr \cr
&=& \left( i \frac{\delta}{\delta J^n_{t_n}} \right)
 \ldots  \left( i \frac{\delta}{\delta J^1_{t_1}}  \right)   \int_{[t_-,t_+]} {\cal D} \phi e^{iS^{AdS}} \
 \langle \beta | \phi(t_+) ;t_+ \rangle
\langle \phi(t_-);t_- | \alpha \rangle.  \ \ \
\end{eqnarray}

Once again the states
 $|\beta \rangle$ and $|\alpha \rangle$ are to be held fixed under the variations in precisely the usual way to be the prescription for the Schwinger variational principle.  Note in particular that this statement refers to eigenstates of the full bulk field operators $\phi(x,t)$ and, as a result, does not require us in any sense to have first split $\phi$ into a ``normalizable" and a ``non-normalizable" part\footnote{Here we use the terminology of \cite{BLK,BLKT}.}.  We now close this section with several remarks.
\FIGURE{\epsfig{file=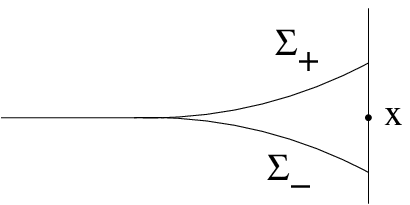}\caption{Two Cauchy surfaces $\Sigma_\pm$ that partly coincide, but respectively provide Cauchy surfaces for the future and past of a point $x$ on the boundary.} \label{touch}}

{\bf Remark 1:} To the extent that the bulk theory may be described as a local field theory, any CFT correlator at time $t_n > t_{n-1} > \ldots > t_1$ can be described as a variation of a path integral performed over any interval $[t_+,t_-]$ with $t_+ > t_n$ and $t_- < t_1$.  In particular, should we consider a semi-classical context with black hole horizons, it is clear that all CFT correlators can be expressed in terms of a path integral over regions of spacetime {\it outside} of these horizons.

{\bf Remark 2:}  Since (\ref{abcors}) is largely independent of the choice of $t_\pm$, we may trivially take the limit in which $t_\pm$ lie to the far future and far past.   In contrast, appendix A shows that, had we attempted to neglect the wavefunctions at $t_\pm$, this limit would in general not be well-defined.

{\bf Remark 3:}  In order to define a time-independent source-free
Hamiltonian $H_0$, we found it convenient above to take the
coordinate $t$ to correspond to an asymptotic time translation.
However, the formalism may be extended to the case where $t$
labels an arbitrary family of Cauchy surfaces $\Sigma_t$. In fact,
under certain conditions (see Fig. \ref{touch}) one make take part
of  $\Sigma_+ = \Sigma_{t_+}$ to coincide with part of $\Sigma_- =
\Sigma_{t_-}$, so that part of the functional integral becomes
trivial.  One sees that in general CFT correlators associated with
boundary points $x_1,\ldots, x_n$ may be computed via a path
integral over any region of spacetime bounded by bulk  surfaces
$\Sigma_\pm$ such that $\Sigma_+$ ($\Sigma_-$) is a Cauchy surface
for the bulk region to the future (past)  of all points $x_i$.
This is essentially the statement that one need integrate only
over the wedge regions described in \cite{R1,R2}.

\section{Semi-classical physics in Lorentz signature}
\label{semi}

We have argued above that for $t_+ \ge t_n \ge \ldots \ge t_1 \ge t_-$ the Lorentzian AdS/CFT correspondence takes the form
\begin{eqnarray}
\label{abcors2}
 \langle \tilde \beta |\tilde {\cal O}_n(t_n)
&\ldots& \tilde  {\cal O}_1(t_1)
| \tilde \alpha\rangle \cr \cr & =& \left( i \frac{\delta}{\delta J^n_{t_n}} \right)
 \ldots  \left( i \frac{\delta}{\delta J^1_{t_1}}  \right)   \int_{[t_-,t_+]} {\cal D} \phi e^{iS^{AdS}} \
 \langle \beta | \phi(t_+) ;t_+ \rangle
\langle \phi(t_-);t_- | \alpha \rangle,  \ \ \
\end{eqnarray}
where $|\tilde \alpha \rangle, |\tilde \beta \rangle$ are arbitrary CFT states with bulk counterparts $|\alpha \rangle, |\beta \rangle$ and where $\tilde {\cal O}_i$ is the CFT operator associated with the variation $\delta J^i$ in the bulk boundary conditions $J =  \{ J^i_t \}$.  All bulk fields are represented by $\phi$ above and the factors on the far right in the path integral are wavefunctions at times $t_\pm$ which are to be held {\it fixed} under variations of the boundary conditions.

We now turn  to the use of semi-classical techniques to calculate the variations (\ref{abcors2}).  We wish in particular to address the role of the wavefunctions at $t_\pm$ that appear in the path integral.  These wavefunctions have not previously received explicit attention though, as shown in Appendix \ref{ambig}, simply ignoring such terms leads to inconsistency.  In practice, such boundary terms are particularly relevant when one wishes to perform various integrations by parts in the classical action.  The general structure of the semi-classical approximation in the presence of such boundary terms is described in  section \ref{gs} below.  This will lead in section \ref{1pts} to the conclusion that all Lorentzian CFT one-point functions are represented by simple boundary terms at null infinity (the usual cylinder boundary) at leading order in the semi-classical expansion about a classical bulk solution.  This result holds even in the presence of black hole horizons.  Finally, we invoke the standard toy model of a bulk scalar field in section \ref{ex} to provide an explicit example in which these issues can be studied in detail.

\subsection{General Structure}
\label{gs}

We consider here the case where {\it all} of the integrations implicit in (\ref{abcors}) may be approximated using semi-classical methods.  Since this includes integrations against the initial and final wavefunctions, this will in particular require that the wavefunctions $\Psi_-(\phi) := \langle \phi; t_- | \alpha \rangle_J$
 and $\Psi_+(\phi) :=  \langle \phi; t_+ | \beta \rangle_J$ be in some sense ``semi-classical."  That is, the state $|\alpha \rangle$ must be semi-classical in the bulk at time $t_-$ and the state $| \beta \rangle$ must be semi-classical in the bulk at time $t_+$.  We note in particular that $|\alpha \rangle, |\beta \rangle$ are {\rm not} required to remain semi-classical for $t \not \in [t_-,t_+]$.

For simplicity, we will take the operators $\phi(t)$ to be Hermitian so that the path integral is over real field histories.

In order to proceed, we write the two wavefunctions in the standard
semi-classical form $\Psi_\pm(\phi) = e^{i\psi_\pm} $.  Note that this introduces no assumptions, and merely defines $\psi_\pm$.  Thus we
may write the innner product of interest  in the form
\begin{equation}
\langle \beta | \alpha \rangle =  \int_{[t_-,t_+]} {\cal D} \phi \
\exp^{i(S^{AdS} + \psi_- - \psi^*_+)},
\end{equation}
where $*$ represents complex conjugation. As a result, it is the object
\begin{equation}
\label{Sab}
S_{\alpha, \beta}[\phi(t)] := S^{AdS}[\phi(t)] +
\psi_-(\phi(t_-)) - \psi^*_+(\phi(t_+))
\end{equation}
which plays the role of
an action for the semi-classical approximation, in the sense that
this approximation picks out field histories about which
$S_{\alpha, \beta}$ is a stationary under variations of $\delta
\phi(t)$ for $t \in [t_-,t_+]$, including variations of
$\phi(t_\pm)$.

However, since we are approximating the
overlap $\langle \beta | \alpha \rangle $ for {\it fixed} $J$, the variations $\delta \phi(t)$ under which $S_{\alpha \beta}$ is stationary are those that preserve the boundary conditions $J$.  Recall that such boundary conditions define the Hamiltonian $H_t$ and thus are associated with the boundaries $\partial\Sigma_t$ of the Cauchy surfaces $\Sigma_t$ defined  by the condition $t=const$.  In particular, boundary conditions specified by $J$ are boundary conditions at null infinity, $I$; i.e., at the cylinder boundary of AdS shown in Fig. \ref{AdSfig}.
 We emphasize once again the different roles of boundary
conditions at $I$ and at $t_\pm$:  one considers only variations
$\delta \phi$ which preserve the boundary conditions at $I$, while
one will require $S_{\alpha \beta}$ to be stationary under
variations that are otherwise arbitrary at $t_\pm$.

Despite our use of real fields, the terms $\psi_\pm$ will typically be complex.  This should not disturb the reader as stationary phase methods in general require one to extend the action (here, $S_{\alpha \beta}$) to complex arguments as an analytic function.  Thus, we need only assume that $\psi_\pm$ admit such analytic extensions.

Neveretheless, one may wonder to what extent $S_{\alpha, \beta}$ may be
expected to have such stationary points.  Recall that, as indicated by the notation in (\ref{Sab}),  both $S_{\alpha, \beta}[\phi(t)]$ and
$S^{AdS}[\phi(t)]$ are functionals which depend on the full
history $\phi(t)$ while $\psi_-(\phi(t_-))$ and
$\psi_+(\phi(t_+))$ depend only on the initial and final values as
indicated.  Thus, $\psi_\pm(\phi(t_\pm))$ may be thought of as
defining explicit boundary terms in the action $S_{\alpha \beta}$ which, although complicated (and
in general quite non-local in space), will not affect variations of
$\phi$ which vanish at $t_\pm$.  Thus, any stationary point will
satisfy the bulk equation of motion for $t_+ > t > t_-$.

Consider now variations about some field configuration that satisfies the bulk equations of motion for
$t_+ > t > t_-$.  When will $S_{\alpha \beta}$ also be stationary with respect to variations $\delta \phi$ which do {\it not} vanish at $t_\pm$?  To answer this question, consider a general such variation of $S^{AdS}$.  Let us in particular consider the case where the Lagrangian is a function only of the fields $\phi$ and their first derivatives\footnote{In general, the presence of second or higher order time derivatives in the Lagrangian means that the phase space cannot be parametrized by configurations and velocities alone.  Instead, the phase space must be extended.  As a result, quantum wavefunctions cannot be specified solely by functions on the configuration space.  The resulting formalism is straightforward, but cumbersome to write explicitly for the general case.  We therefore choose to keep the notation simple and to assume that any such Lagrangian has been reformulated as a function only of configuration and velocity degrees of freedom, possibly through the introduction of a sufficiently large number of additional fields.}, so that that quantum states are fully specified by functions of the configuration fields.
We rely on the details of the skeletonization procedure which, as noted previously, defines an action on phase space of the standard form (\ref{action}), whose variation is then
\begin{equation}
\label{vps} \delta S = \int_{t \in [t_-,t_+]}
(\delta p \dot{q} - \dot{p} q - \delta H_{J^t}(t)) \ dt +
\int_{\Sigma_{t_+}} p (t_+)\delta q({t_+}) - \int_{\Sigma_{t_-}}
p(t_-) \delta q(t_-).
\end{equation}
Here $\Sigma_{t}$ are the surfaces in AdS associated with constant values of the time
$t$ and the various terms are understood to include both integrals
over position and a sum over whatever fields are to be varied; in particular, we have suppressed the spatial volume elements in each of the three terms above.

Let us pause to comment on the general form of $\delta
H_{J^t}(t)$, which implicitly involves an integral over space. Locality tells us that
this variation takes the form
\begin{equation}
\delta H_{J^t}(t) = \int_{\Sigma_t} \left( \frac{\delta H}{\delta
q} \delta q + \frac{\delta H}{\delta p} \delta p \right) +
\int_{\partial \Sigma_t} b_{J^t}(\delta q, \delta p),
\end{equation}
where $b$ is an appropriate boundary term which will in general
depend on the boundary conditions $J^t$. We shall assume that the
boundary term $b_{J^t}$ vanishes\footnote{This is naturally taken
to be the condition which determines any explicit dependence of
$S^{AdS}$ on the boundary conditions $J$, since it is equivalent
to the requirement that action is indeed stationary when the bulk
equations of motion are satisfied.  For the simplest boundary
conditions, it follows in cases of relevance to AdS/CFT  from
\cite{Witten,BLK,BLKT}.  As noted in \cite{PM04} and described in
section \ref{ex} below, the ``improved action'' of \cite{KW}
(their equation (2.14)) also satisfies this condition for certain
more general boundary conditions.  The same is true of the actions
in \cite{HH} for the boundary conditions considered there.} for
all variations $(\delta q, \delta p)$ which preserve the boundary
conditions $J^t$.   Note that since $\Sigma_t$ is a Cauchy
surface, its boundary $\partial \Sigma_t$ is a cross-section of
null infinity $I$.

Since the process of forming the second order action $S^{AdS}$
from the first order action (\ref{action}) simply involves
inserting the appropriate function $p(q,\dot{q})$ for the momenta
$p$, it is clear that the variation of $S^{AdS}$ is obtained from
(\ref{vps}) by the same rule.   The result is
\begin{eqnarray}
\label{varyAdS} \delta S^{AdS} &=& \int_{(t_-,t_+)} \frac{\delta
S^{AdS}}{\delta \phi(t)} \delta \phi(t) + \int_{I}
b_{J^t} (\delta q, \delta p) \cr &+&
 \int_{\Sigma_{t_+}}
p(\phi(t_+),\dot \phi(t_+) )  \delta \phi(t_+) -
\int_{\Sigma_{t_-}} p(\phi(t_-),\dot \phi(t_-))  \delta \phi(t_-),
\end{eqnarray}
where $\frac{\delta S^{AdS}}{\delta \phi(t)} =0$ are the bulk
equations of motion for $t_+ > t > t_-$. An important point is
that one sees explicitly that the variations of the velocities $\delta \dot \phi(t_\pm)$
will not appear in boundary terms\footnote{One may
in any case have suspected this from the idea that the action $S^{AdS}$
defines a path integral which, if one does not integrate over the
boundary values $\phi(t_\pm)$ of the fields, is designed to
calculate the overlap $\langle \phi_+ ; t_+ | \phi_- ; t_-
\rangle$ between field eigenstates. Such field eigenstates clearly
fix the boundary values to be $\phi(t_\pm) = \phi_\pm$, and the
action should in the classical limit yield a well-defined
variational problem with the corresponding boundary conditions
$\delta \phi (t_\pm)=0$. This would not be the case if the above
boundary terms contained variations $\delta \dot \phi (t_\pm)$ of
the velocities.} at $\Sigma_{t_\pm}$.

Since $b_{J^t}(t)$ vanishes under variations that preserve the
boundary conditions $J$,  the full action $S_{\alpha \beta}$ will
be stationary under such variations when
\begin{eqnarray}
\label{stat}
 \frac{\delta S^{AdS}}{\delta \phi(t)} = 0 \ &{\rm for
}& \ t_+
> t > t_-,  \cr    p(\phi(t_-),\dot \phi(t_-))
= \frac{\delta \psi_-}{\delta \phi(t_-)}\ \ \ &{\rm and }& \ \ \
p(\phi(t_+),\dot \phi(t_+)) = \frac{\delta \psi_+^*}{\delta
\phi(t_+)}.
\end{eqnarray}
While detailed analysis is required in order to determine the
existence and uniqueness of such stationary points, one sees that
we have the usual sort of boundary value problem that one expects
to obtain from a variational principle: Stationary points
correspond to histories $\phi(t)$ which satisfy both  the bulk equations
of motion and  boundary conditions which, at each end, are
determined by one {\it complex} relation for each field between
configuration and velocity variables.  Because, as remarked above,
stationary points are naturally sought in the space of complex
solutions $\phi(t)$, one does indeed have the appropriate setting
for a semi-classical formalism.  Note in particular that there is no freedom to add an arbitrary solution to the bulk equations of motion.  We will see in an example below
the important role played by this use of complex fields in, for
example, obtaining the usual Feynmann two-point function (whose
origin may seem obscure in this formalism) in the
case where $|\alpha \rangle$ and $|\beta \rangle$ are vacuum
states of some time-translation-invariant Hamiltonian.

\subsection{One-point functions}

\label{1pts}

We have seen that the semi-classical approximation to $\langle \alpha | \beta \rangle$ with boundary conditions $J$ picks out a particular (complex) history, which we may call $\phi_{J,\alpha, \beta}(t)$.  This history is a stationary point of $S_{\alpha \beta}$ under all variations $\delta \phi (t)$ for $t \in [t_-,t_+]$ such that $\delta \phi(t)$ preserves the particular boundary conditions $J$ at null infinity.  At leading order in the semi-classical approximation we have
\begin{equation}
\label{semiA} \langle \alpha | \beta \rangle_J = \exp(i S_{\alpha,
\beta} [ \phi_{J,\alpha, \beta} ] ),
\end{equation}
where we have now  added the subscript $J$ to the left-hand side
to remind the reader that, because $|\alpha \rangle$ and $|\beta
\rangle$ satisfy respectively retarded and advanced conditions,
this inner product does indeed depend on the boundary conditions
$J$ at null infinity.

One may proceed to calculate any CFT $n$-point function in this
approximation through variations of $S_{\alpha, \beta} [
\phi_{J,\alpha, \beta} ]$ with respect to $J$.  Let us consider
the particular case of a one-point function, which corresponds to
a first variation.  Since we have already computed the
first variation of $S_{\alpha \beta}$ in (\ref{varyAdS}), computation of our one-point function merely requires substitution of $\delta \phi = \frac{\delta \phi_{J, \alpha, \beta}}{\delta J} \delta J$ and evaluation of the result on $\phi_{J,\alpha, \beta}$.  Since by construction  $\phi_{J,\alpha, \beta}$
satisfies (\ref{stat}), we have

\begin{equation}
\label{1pt} \delta S_{\alpha \beta}[ \phi_{J,\alpha ,
\beta} ]
 = \int_{I}   b_{J^t}\left( \frac{\delta \phi_{J,\alpha \beta}}{\delta J} \delta J\right),
 \end{equation}
where the right-hand side is the boundary term $b_{J^t}$ evaluated on the variation
$\delta \phi = \frac{\delta \phi_{J, \alpha, \beta}}{\delta J} \delta J$.
Thus, a generic CFT one-point function is determined at leading
order in the semi-classical limit by a boundary term at null
infinity.

Finally, it is appropriate to comment on the observation
above that, in general, the boundary conditions at $t_\pm$ in
(\ref{stat}) will require the stationary point $\phi_{J,\alpha,
\beta}$ to be complex, even though the corresponding $\phi(t)$
were taken to be Hermitian.  To clarify this point, consider the
special case for which i) $|\alpha \rangle = |\beta \rangle$, ii) at each
time $t \in [t_-,t_+]$ the wavefunction $\langle \phi; t| \alpha
\rangle$ is sharply peaked about a real classical solution
$\phi_{J, \alpha, \alpha} (t)$, and iii) for which the corresponding
wavefunction in momentum space is sharply peaked about the
momentum corresponding to the solution $\phi_{J, \alpha, \alpha}(t)$. Then the
real solution $\phi_{J, \alpha, \alpha}(t)$ will indeed satisfy
(\ref{stat}), as $-i \frac{\delta }{\delta \phi}\Psi_\pm$ gives
the action of the momentum operator on the wavefunction at
$t_\pm$.  Thus, we find that the stationary phase solution is indeed
real in the case where $|\alpha \rangle = |\beta \rangle$ and the
state is semi-classical in the usual sense over the time interval
$t \in [t_-,t_+]$.  In this case the one-point functions (\ref{1pt}) are
also real, in accord with the statement that $\phi(t)$ are
Hermitian.  Nevertheless, complex solutions can still become
relevant when second and higher variations are computed.  We shall
see how this works in more detail in the example below.  There
such considerations lead to the usual Feynmann propagator when
$|\alpha \rangle$ is taken to be a vacuum state.

\subsection{An illustrative example: The scalar test field in detail}

\label{ex}

At this point, the reader may feel that an illustrative example is
desperately needed in order to make more concrete the  rather
general and abstract considerations above.  Let us therefore consider
the usual toy model in which the bulk theory is replaced by a
real scalar test field $\phi$ in AdS${}_{d+1}$.  Here for simplicitly we set the AdS length scale $\ell$ to be $\ell = 1$.  We use coordinates such that the AdS${}_{d+1}$ metric
is
\begin{equation}
ds^2 = g_{ab} dy^a dy^b =  - (1 + r^2)dt^2 + \frac{dr^2}{1+ r^2} + r^2 d\Omega_{d-1}^2,
\end{equation}
where $d\Omega_{d-1}^2$ is the round metric on the unit $S^{d-1}$.

The boundary conditions $J$ are taken to parametrize various
possible asymptotic behaviors of $\phi$ near null infinity.    Suppose that our scalar is associated with a potential $V(\phi)$ with squared mass $m^2 = \frac{1}{2} V''(0)$.  Then for $0 \ge m^2 >  -d/2$, one finds that all solutions to the equations of motion take the asymptotic form
\begin{equation}
\label{asympt}
\phi \rightarrow \frac{a(x)}{r^{\lambda_-}} + \frac{b(x)}{r^{\lambda_+}},
\end{equation}
where $x$ are coordinates on null infinity ($I$) and where
\begin{equation}
\lambda_\pm = \frac{d}{2} \pm \frac{1}{2} \sqrt{d^2 + 4 m^2}.
\end{equation}
This asymptotic form also holds for $m^2 > 0$ if the potential is
purely quadratic\footnote{However, for $m^2 > 0$ we have
$\lambda_-< 0$.  As a result, one solution to the linearized
equation grows near infinity  and non-linear terms can have a
significant effect.}.  Let us therefore assume that either $0 \ge
m^2 >   - d/2$ or  $m^2 > 0$ with $V(\phi) = \frac{1}{2} m^2 \phi^2$
so that we may use the behavior (\ref{asympt}).  Note that for
simplicity we have forbidden our scalar from saturating the
Breitenlohner-Freedman bound \cite{BF}.

Consider the action
\begin{equation}
\label{KWA}
S^{AdS} = - \int_{t\in [t_-,t_+]} \left( \frac{1}{2}\partial
\phi^2 + V(\phi) \right) \sqrt{-g} - \frac{1}{2} \lambda_- \int_I \sqrt{-g_I} \phi^2.
\end{equation}

As noted in \cite{PM04}, this action is equivalent to the
``improved action" advocated by Klebanov and Witten (see equation
(2.14) of \cite{KW}) for configurations satisfying (\ref{asympt}).
Here $g_I$ denotes the induced metric on null infinity.  Both
terms diverge for configurations satisfying (\ref{asympt}), but
the particular combination (\ref{KWA}) can be defined by the usual
procedure of regulating the action by moving the boundary to a
finite location. For the full action (\ref{KWA}), the limit where
the boundary is taken to null infinity converges.

An interesting case is where one requires the boundary condition
\begin{equation}
\label{aisJ}
a(x) = J(x), \ \ \ {\rm for} \  x \in I,
\end{equation}
such that one fixes the behavior of the more slowly decreasing term in (\ref{asympt}).  For a solution, the coefficient $b(x)$ is then to be determined by the equations of motion and the initial conditions.
Under the boundary conditions (\ref{aisJ}) we wish to check that
(\ref{KWA}) leads to well-defined equations of motion.  The variation
of $S_{AdS}$ is
\begin{eqnarray}
\label{varyex} \delta S^{AdS} &=&  \int_{t\in [t_-,t_+]} \sqrt{-g} \left( \nabla^2
\phi - V'(\phi)\right)  \delta \phi - \sum_\pm \int_{\Sigma_{t_\pm}}
\sqrt{g_{\Sigma_{t_\pm}}} (n_{\Sigma_{t_\pm}}^a
\partial_a \phi) \delta \phi \cr &-& \int_I  \sqrt{-g_I} (n_I^a
\partial_a \phi) \delta \phi  - \lambda_- \int_I \sqrt{-g_I} \phi \delta \phi,
\end{eqnarray}
where $n_I, n_{\Sigma_{\pm}}$ are outward pointing unit normals (i.e., with $n^a n^b g_{ab} = \pm1$).
To compute the classical bulk equations of motion we need only consider variations with $\delta
\phi(t_\pm) =0$, so that the final two boundary terms vanish.  However, the boundary terms at null infinity must be treated with more care.

Since the above variation occurs at fixed $J$ and respects the
boundary condition $a(x) = J(x)$, we have $\delta a = 0$.  Using
this fact and the asymptotic behavior (\ref{asympt}) it is
straightforward to show that the boundary terms at null infinity
do indeed cancel\footnote{In fact, for the boundary conditions
(\ref{aisJ})  any local boundary term at null infinity built from
the metric, $\phi$, and derivatives of $\phi$ (either along or
transverse to the boundary) is equivalent to the one used in
(\ref{KWA}) if it leads to 1) a finite action and 2) an action
which is stationary when the bulk equations of motion are
satisfied.  This observation may be used \cite{PM02,PM04} to
further justify the choice of boundary terms made in \cite{KW}.
Related comments also appear in \cite{PM01}.}. Thus, for
variations of this form we find
\begin{equation}
\label{varyex2} \delta S^{AdS} =  \int_{t\in [t_-,t_+]} \sqrt{-g} \left( \nabla^2
\phi - V'(\phi)\right)  \delta \phi,
\end{equation}
so that the action is indeed stationary when the bulk equations of motion are satisfied.

Let us now consider states $|\alpha \rangle$, $|\beta \rangle$
which are Gaussian at time $t_-,t_+$ respectively.   When $\phi$
is free, this family of states includes the vacuum state
$|0\rangle$.  More generally, the vacuum formally becomes Gaussian as $\hbar \rightarrow 0$ and one may attempt to construct the vacuum perturbatively.   We will not explore such
perturbation theory in detail here, but it may be interesting to do so in order to obtain a fully Lorentzian formulation of the problem.

We therefore suppose that we have wavefunctions of the form
\begin{eqnarray}
\label{Gauss} \langle \phi; t_- | \alpha \rangle &=& N_+ \exp(i
\phi \pi_-) \exp[-\frac{1}{2}(\phi - \phi_-)C_+(\phi - \phi_-)], \cr \langle
\phi; t_+ | \beta \rangle &=& N_- \exp(i \phi \pi_+) \exp[-\frac{1}{2}(\phi -
\phi_+)C_-(\phi - \phi_+)],
\end{eqnarray}
where $(\phi_\pm, \pi_\pm)$ are points in the phase space
associated with times $t_\pm$ and satisfying the appropriate
boundary conditions set by $J^{t_\pm}$.  Note in particular that
the choice of boundary conditions $J^t$ for $t\in (t_-,t_+)$ has
no bearing on the choice of $(\phi_\pm, \pi_\pm)$.  The operators
$C_\pm$ are to be appropriate positive-definite self-adjoint
linear operators which are similarly compatible with the
boundary conditions $J^{t^\pm}$ and which are independent of $J^t$ for $t\in
(t_-,t_+)$.  In particular, if $|\alpha \rangle$ is the vacuum state of a (stable) linear theory, then $C_+$ is just the frequency operator $\omega$.   Lastly, $N_\pm$ are formal normalization
coefficients\footnote{Divergences in $N_\pm$ are of course more
rigorously dealt with by expressing $|\alpha\rangle$ and $|\beta
\rangle$ as more abstract Gaussian measures with covariance determined by
$C_\pm$.} associated with the determinants of $C_\pm$.

We now turn to the evaluation of $\langle \beta | \alpha \rangle_J$ in the
stationary phase approximation.  The result will be of the form
(\ref{semiA}), where $\phi_{J, \alpha, \beta}$ satisfies the bulk
equations of motion, the boundary conditions at null infinity,  and the conditions
\begin{equation}
\label{pmbc}
\pm \sqrt{g_{\Sigma_{t_\pm}}} n^a \partial_a
\phi_{J, \alpha, \beta} = \pi_\pm \mp i C_\pm (\phi_{J,
\alpha, \beta}(t_\pm) -  \phi_\pm),
\end{equation}
at $t =t_\pm$.  In particular, $\phi_{J, \alpha, \beta}$ will be a
solution if it agrees with $\phi_\pm$  and if its momentum agrees
with $\pi_\pm$.  Note that the condition that $\phi_{J, \alpha,
\beta}$ be real is therefore just the condition that that $\phi_\pm,
\pi_\pm$ be chosen such that a real classical solution connects
these points in phase space over the time interval $[t_-,t_+]$; i.e., that
at least at the semi-classical level our wavefunctions at $t=t_\pm$
are related by time evolution (so that $|\alpha \rangle \approx
|\beta \rangle$ at this level).

Computations of the CFT $n$-point functions will involve the functional derivative
\begin{equation}
K_{J,\alpha,\beta}(y,x) := \frac{\delta \phi_{J,\alpha,\beta}(y)}{\delta J(x)}.
\end{equation}
Note that $K_{J,\alpha, \beta}$ satisfies the bulk equations of motion linearized about $\phi_{J, \alpha, \beta}$ as well as the condition
\begin{equation}
\label{Klim}
\lim_{y \rightarrow x} r_y^{\lambda_-}  K_{J,\alpha,\beta} (y,x') = \delta_I(x,x') \ \ \ {\rm for} \  x,x' \ {\rm on} \  I,
\end{equation}
where $r_y$ is the $r$-coordinate of the point $y$ and
$\delta_I(x,x')$ is the delta-function on $I$ which is a density
with respect to $x'$.  As a result, $K_{J,\alpha,\beta}$ is a
``bulk-to-boundary propagator" in the sense of \cite{BLK,BLKT}. As
remarked in these references, there are many such propagators as
the above conditions allow one to add any solution to the
linearized equations of motion satisfying trivial boundary
conditions at $I$.  Nonetheless, variation of the conditions
(\ref{pmbc}) shows that $K_{J,\alpha,\beta}$ is the (unique) such
propagator satisfying
\begin{equation}
\label{Kbound} \sqrt{g_{\Sigma_{t_\pm}}} (n_{\Sigma_{t_\pm}}^a \frac{\partial}{\partial
y^a} + i  C_\pm ) K_{J,\alpha, \beta}(x,y) = 0
\end{equation}

at times $t_\pm$.

Using (\ref{varyex}), (\ref{Kbound}), and the fact that $\phi_{J, \alpha , \beta}$ satisfies the bulk equation of motion, one finds that
the one-point function is
\begin{eqnarray}
\label{1ex} i \frac{\delta }{\delta J(x)} \langle \beta | \alpha
\rangle_J &=& e^{i\left( S_{\alpha \beta}[\phi_{J, \alpha,
\beta}]\right)}\int_I dx' \sqrt{-g_I} \left( n_I^a
\partial_a \phi_{J, \alpha, \beta}(x')  + \lambda_- \phi_{J, \alpha, \beta} \right) K_J(x,x') \cr
&=& e^{i\left( S_{\alpha \beta}[\phi_{J, \alpha,
\beta}]\right)}  (\lambda_- - \lambda_+) b_{J,\alpha,\beta}(x),
\end{eqnarray}

where asymptotically we have
\begin{equation}
\phi_{J,\alpha,\beta} \rightarrow \frac{a_{J, \alpha, \beta}(x)}{r^{\lambda_-}} +  \frac{b_{J, \alpha, \beta}(x)}{r^{\lambda_+}}.
\end{equation}
The terms involving $a_{J,\alpha, \beta}(x)$ have cancelled due to
the particular choice of boundary term in (\ref{KWA}).  The result
normalizes the one-point functions in the manner advocated in \
\cite{KW,FMMR,G,LMRS}.

Let us now consider the connected two-point function.  For $x_1 \neq x_2$ we obtain
\begin{eqnarray}
\label{21}
- \frac{\delta^2 }{\delta J(x_1) J(x_2)} \ln \langle
\beta | \alpha \rangle_J &=& - i \frac{\delta^2 }{\delta J(x_1)
J(x_2)} S_{\alpha \beta}[\phi_{J,\alpha, \beta}] \cr
&=&  + i  \int_I dx \sqrt{-g_I} \left[n_I^a
\partial_a K_{J,\alpha, \beta}(x_2,x) + \lambda_- K_{J,\alpha,\beta}(x_2,x) \right]K_{J,\alpha,
\beta}(x_1,x) \cr  &=& +i  (\lambda_- - \lambda_+) \lim_{y \rightarrow x_1}  r_y^{\lambda_+}  K_{J,\alpha,
\beta}(x_2,y),
\end{eqnarray}
where in the second step we have used the fact that $K_{J,\alpha, \beta}$ satisfies the linearized equations of motion about $\phi_{J,\alpha, \beta}$.
Thus, the CFT two-point function is associated with the limiting form of the particular bulk-boundary propagator $K_{J,\alpha,\beta}$ which is directly determined by the states $|\alpha \rangle, |\beta \rangle$ through the conditions (\ref{Kbound}).   In the particular
case above where $|\alpha \rangle, |\beta \rangle$ do happen to be vacuua, the operator $C_\pm$ is the the frequency operator $\omega$ so that
$K_{J,\alpha, \beta}$ is precisely the Feynmann propagator
associated with these vacuua.   As a result, our prescription agrees with that of \cite{CS} in the context studied there.

We have written (\ref{21}) as a boundary term, though even in the
case where $|\alpha \rangle = |\beta \rangle$ and where we may
approximate the state as sharply peaked around a classical
solution $\phi(t)$ this result will in general depend on the
values of $\phi_{J,\alpha,\beta}$ in the interior of the
spacetime.  This is a natural consequence of specifying $K_{J,\alpha,
\beta}$ in the interior as the solution to a particular boundary
value problem whose bulk equations of motion are determined by
linearizing about the semi-classical solution
$\phi_{J,\alpha,\alpha}$.

\section{Discussion}
\label{disc}

We have considered various subtleties of the Lorentzian formulation of the
AdS/CFT correspondence, and in particular the specification of how CFT $n$-point functions may
be computed from variations of bulk path integrals with respect to
boundary conditions.  Though our interest was in $n$-point functions associated with non-vacuum states in the Lorentzian theory, our strategy was to carefully derive the Lorentzian correspondence
via analytic continuation from the Euclidean, where the details of
the dictionary (at the level discussed here) are more clearly specified
in the literature and which is inherently free of issues
associated with the choice of propagating states.  We have made an
effort to be pedagogical as, while the various steps involved are
familiar, there are many potential subtleties in applying them to
the AdS/CFT context.  We hope that our pedagogical presentation
has made all such issues transparent.

The main result is that
at leading order in the semi-classical approximation  the AdS/CFT correspondence takes the form
\begin{equation}
\label{mainr}
 \langle \tilde \alpha | \tilde \beta \rangle_{J} = \exp(i S_{\alpha,
\beta} [ \phi_{J,\alpha, \beta} ] ),
\end{equation}

where $|\tilde \alpha \rangle, |\tilde \beta\rangle$ are arbitrary CFT states while $S_{\alpha  \beta}$
is given by (\ref{Sab}) and is formed from the AdS bulk action $S^{AdS}$ together with additional (complex) boundary terms at times $t_\pm$ associated with the bulk wavefunctions at $t_\pm$ corresponding to $|\tilde \alpha \rangle, |\tilde \beta\rangle$.  This action is evaluated on the particular history $\phi_{J, \alpha, \beta}$ which is a (perhaps complex) stationary point of $S_{\alpha \beta}$ and satisfies a set of bulk boundary conditions specified by $J$.  As usual, $J$ also specifies a set of sources in the CFT, and the associated dependence of  the inner product on the left of (\ref{mainr}) is indicated by the subscript $J$.

Expression (\ref{mainr})  is, as it stands,  independent of the choice of $t_+ , t_-$, so long as $t_+ \ge t_-$.  This is manifestly so on the left-hand side, and occurs on the right because the semi-classical evolution of the wavefunctions is determined directly by the  AdS bulk action $S^{AdS}$ in such a way that the full $S_{\alpha \beta}$ remains invariant.  However, in order to compute CFT correlators by varying the parameters $J$, one requires a notion of what is the ``same'' state (e.g., $|\alpha \rangle$) in systems with two distinct values of $J$.  The correct notion is that $|\tilde \alpha \rangle$ and the corresponding bulk wavefunction must be held fixed in the sense of retarded boundary conditions, while $|\tilde \beta \rangle$ and the corresponding bulk wavefunction must be held fixed in the sense of advanced boundary conditions.
Thus, it is only natural to use (\ref{mainr}) in the case where such variations are restricted to times $t$ between $t_+$ and $t_-$.  As a result, the simplest case occurs when $|\beta \rangle$ may be considered to be defined at $t_+$ while $|\alpha \rangle$ is defined at $t_-$, so that we naturally have what is often called an "in-out" matrix element.
The result is that the AdS/CFT dictionary is naturally expressed as an implementation of the Schwinger variational principle \cite{Schwing,Bryce}.

Our careful study of the boundary terms at $t_\pm$ has resulted in
certain differences from or clarifications of the prescription
suggested in \cite{GKP,BLKT}.      In brief, these features are

\begin{enumerate}
\item CFT correlators associated with boundary points $x_1,\ldots,
x_n$ may be computed via a path integral over any region of
spacetime bounded by bulk  surfaces $\Sigma_\pm$ such that
$\Sigma_+$ ($\Sigma_-$) is a Cauchy surface for the bulk region to
the future (past)  of all points $x_i$.  In other words, the path
integral refers only to the wedge regions described in
\cite{R1,R2}.

\item  As a result of (1), all CFT correlators can be expressed in terms of a path integral over regions of spacetime {\it outside} of any black hole horizons\footnote{Here we have in mind the usual setting of an asymptotically AdS bulk with a single asymptotic region.  However, the principle may be generalized to situations having two or more asymptotic regions separated by bifurcate horizons, and to correlators relating operators in different such regions.  In that case, one may deform both the initial and final surface to pass through the bifurcation surface(s).  As a result, one may express the correlator in terms of a bulk path integral over regions that do not contain trapped surfaces and which, in this sense are  again "outside" the black hole.}

\item  Also as a result of (1), the expression $ S_{\alpha,
\beta} [ \phi_{J,\alpha, \beta} ] $ has a well-defined limit as $t_\pm \rightarrow \infty$.  As described in the appendix, this is in contrast to the prescription of \cite{BLKT}.

\item \label{det} The semi-classical solution $ \phi_{J,\alpha, \beta}$ is determined entirely by $S_{\alpha \beta}$ together with the boundary conditions $J$. Thus, as one might expect from CFT considerations, the appropriate ``bulk/boundary propagator" is also determined directly by the quantum states $|\alpha\rangle, |\beta \rangle.$
There is no freedom to add an arbitrary solution to the bulk equations of motion corresponding to a separate choice of ``vacuum'' state.

\item  The boundary terms in $S_{\alpha \beta}$ interact with result (\ref{det}) in just such a way that,
 at  leading order in this approximation, any CFT one-point function is given by a simple boundary value of the classical bulk solution at null infinity, $I$.  This result holds even in the presence of black hole horizons.
\end{enumerate}

 We have exhibited such features in detail using a common
toy model involving a scalar test field.
This toy model can be obtained as a limit of the full AdS/CFT
correspondence where one is interested in a bulk scalar field and
when one can ignore the interactions with other bulk fields.  It
is clear that the general case is similar, at least in the approximation that one expands about a classical solution.

In general, the above formalism is useful when one has chosen
states $|\alpha \rangle, |\beta \rangle$ whose bulk wavefunctions
can be determined at times $t_\pm$.  Note that this may be
non-trivial in an interacting theory, for example, if one wishes
to compute correlation functions in some vacuum state $|0\rangle$.
Thus, as usual, in this case it may be more efficient to compute
correlators via analytic continuation from the Euclidean, as this
will result in the appropriate correlators for $|0\rangle$ and as
the wavefunction of $|0\rangle$ will {\it not} be needed for the
Euclidean calculation.  This is the context considered
by \cite{KOS,LS}, which showed how the Lorentzian wavefunction could be written as a path integral over a complex contour.

The formalism described in the present work is precisely adapted
to the sort of question posed in \cite{BL,BR}, in which one
explores $n$-point functions in a CFT state dual to a non-trivial
propagating bulk solution.  The situation explored in \cite{BL,BR}
was particularly interesting, as it involved a classical solution
in which a wave packet $\phi_{wp}$ of a scalar field falls into a
rotating black hole. A portion of the conformal diagram for such
black holes is shown in Fig. \ref{RBTZ}, which indicates the
presence of a Cauchy horizon inside the usual event horizon of the
black hole.  As one expects from the study of similar black hole
solutions \cite{PI,Ori,Dafermos}, this Cauchy horizon is unstable
and the solution $\phi_{wp}$ has a stress-energy tensor which
generically diverges at the inner horizon.  The question  in
\cite{BL,BR} was whether any CFT $n$-point function could be
sensitive to this divergence, and the issue was explored using the
bulk semi-classical approximation together with proposals for the
Lorentzian AdS/CFT dictionary based on reasonable-sounding
extrapolations of the results of \cite{KOS,LS} for the vacuum
case.

\FIGURE{\epsfig{file=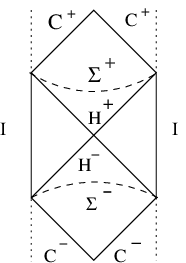, width=1.5in}\caption{The conformal diagram for a rotating BTZ black hole.  We have labeled the  future and past Cauchy horizons $C^\pm$, the future and past event horizons $H^\pm$, null infinity $I$, and surfaces $\Sigma_\pm$ (dashed arcs) which lie to the future (past) of $I$ but are otherwise arbitrary.  The singularities are the the vertical dotted lines.} \label{RBTZ}}

Now, the perturbed black hole represents a non-vacuum state in the
usual AdSaaaaaa/CFT Hilbert space. Having carefully developed the
Lorentzian dictionary for such settings above, result (1) above
tells us that any such $n$-point function can be computed from a
path integral which integrates only over the region between two
{\it arbitrarily chosen} Cauchy surfaces $\Sigma_{\pm}$ which lie
to the future and past of all of null infinity.  Some examples are
shown in Fig. \ref{RBTZ}.   Since \cite{BL} shows that $\phi$
 diverges only on the Cauchy horizons,  we see that the instability cannot affect the CFT
$n$-point functions. In fact, result (1) shows that, at the level
discussed in this work, the AdS/CFT dictionary does not allow CFT
correlators to peer inside black holes any more than operators
near infinity in a local field theory are sensitive to black hole
interiors\footnote{One may point out that black hole interiors are
also determined, via the equations of motion, by the exterior
region.  Thus, there is a sense in which information inside a
black hole {\it is} accessible to a local field theory in the
exterior. This point, however, seems to merely avoid asking the
interesting questions about black holes.}.  It is important to
keep this in mind when considering the implications of results
(e.g., as in \cite{KOS,LS,FHKS}) obtained in {\it analytic}
spacetimes which relate correlators to black hole interiors.

Note that we do not rule out the possibility that a more sophisticated treatment of AdS/CFT may in fact endow CFT correlators with such insight.  Because the AdS theory is a string theory (and thus a theory of quantum gravity), the manipulations above are largely formal.  It remains to be seen to what extent they are fully justified.  Nevertheless, we emphasize that they should be justified to the extent that the bulk theory may be approximated by expanding about a classical bulk solution as is the case in most treatments to date.
In the most optimistic case, one might perhaps move beyond this approximation through a  semi-classical analysis of the
gravitational field.  For example, consideration of complex metrics naturally allows for
changes of spacetime signature, and thus the possibility of
spacetime topology change.  Such topology change departs from the
local field theory behavior assumed in our analysis above, as it
is not obviously associated with a local Hamiltonian.  As a
result, we are sympathetic to attempts such as \cite{J2,SH} to use
topology and signature changing metrics to probe questions
concerning black hole information.

\medskip

{\bf Acknowledgments:}  This work is dedicated to the memory of Bryce S. DeWitt, and to his influence on both theoretical physics in general and on the author in particular.  The remarks in this paper largely study the relation of the AdS/CFT dictionary to the Schwinger variational principle, which the author first learned from DeWitt as a Ph.D. student.
The author would also like to thank Vijay Balasubramanian, Thomas Levi, and Simon Ross for numerous lengthy discussions, and  Marcus Berg, Michael Haack, Thomas Hertog, Stefan Hollands, Per Kraus, and Mark Srednicki for a number of useful comments.  This work was supported in part by NSF grant PHY0354978 and by funds from the  University of California.

\appendix

\section{Semi-classical ambiguities}

\label{ambig}

The full form of the Lorentzian AdS path integral was derived in section \ref{map} above in terms of an action $S_{\alpha \beta}$ associated with a time interval $[t_-,t_+]$ and which includes certain boundary terms at $t_\pm$.  Here we emphasize the key role played by such boundary terms by considering the effect of simply dropping such terms on the toy model of section (\ref{ex}).  That is, we consider the effect of replacing $S_{\alpha \beta}$ with $S^{AdS}$ as would occur if one strictly follows the previous literature \cite{BLK,BLKT}.  In this context, we will refer to the resulting path integral as the Lorentzian partition function $Z_J$.
 We will in particular be interested in the limit of $Z_J$ in which $t_\pm$ are taken to $\pm \infty$, as in that case one might hope that the contribution of boundary terms could be ignored.

As in section (\ref{ex}), we consider a semi-classical setting.  The standard assumption is then that one may approximate
\begin{eqnarray}
\label{oldrule} Z_J = exp ( iS^{AdS}[\phi_{J}]), \ {\rm
where} \ \cr S^{AdS}[\phi] =-  \int \sqrt{-g} \left( \frac{1}{2}
\partial_a \phi \partial^a \phi + V(\phi) \right) - \frac{1}{2} \lambda_- \int_I \sqrt{-g_I} \phi^2,
\end{eqnarray}
where the details of the boundary term at null infinity ($I$) were defined in section (\ref{ex}) and $\phi_J$ is a stationary point of $S^{AdS}$ up to boundary terms in the far past and future.
Note that, in contrast to the treatment in section (\ref{ex}), the semi-classical solution $\phi_J$ is now {\it not} fully specified by the above requirement that $S^{AdS}$ be stationary.  One knows only that $\phi_J$ is a solution to the bulk classical equations satisfying the boundary conditions $J$ at null infinity. The literature assumes that one works near some particular classical solution $\phi_0$, which for convenience we have taken to be associated with the boundary conditions $J=0$, and takes $\phi_J$ to be of the form

\begin{equation}
\phi_{J}(y) = \phi_0(y) + \int_I K_J(x,y) J(x),
\end{equation}

where  $K_J(x,y)$ is such that  $\phi_J(y)$ again solves the equations of motion and the normalization condition (\ref{Klim}).  Here we have written an expression appropriate for non-linear theory, but taking $J=0$ yields a `propagator' $K_0(x,y)$ which satisfies the linearized equations of motion about $\phi_0$.
For example, following \cite{BL}
(and inspired by \cite{KOS,LS}), one might take $K_0(x,y)$ to be some Feynmann-like propagator.

Variations of  $S_{AdS}[\phi_{J}]$ with respect to $J$
are to generate the CFT $n$-point functions.  Unfortunately, the results are not well-defined.  This is so even for the zero-point function $Z_J$ itself.  Consider for example a free field with $V(\phi)=\frac{1}{2}m^2 \phi$ and any case in which $J$ becomes trivial to the far future and far past but  the solution $\phi_0$ of interest does not.  Then $S^{AdS}[\phi_J]$ is most readily evaluated after an integration by parts, though since we want the numerical value of $S^{AdS}[\phi_J]$ we cannot discard the boundary term.
  We have

\begin{equation}
\label{problem} S[\phi_{J}] = S_0 + \lim_{t_+
\rightarrow + \infty} \int_{\Sigma_{t_+}}
\sqrt{g_{\Sigma_f}}  \phi_J n_{\Sigma_{t_+}}^a \partial_a \phi_J +
\lim_{t_- \rightarrow - \infty} \int_{\Sigma_{t_-}}
\sqrt{g_{\Sigma_{t_-}}} \phi n_{\Sigma_{t_-}}^a \partial_a \phi_J,
\end{equation}
where $S_0$ is a finite contribution from the region of $I$ in which $\phi_0$ is nontrivial.  Note that in the current (free field) case the remaining bulk integral vanishes by the equations of motion.
In order for (\ref{problem}) to converge, the boundary terms must each approach a well-defined value in the limit $t_\pm \rightarrow \pm \infty$.  But they do not.  Instead, such terms are quasi-periodic in time, as are all quantities computed from free fields in AdS space.

The same issue arises when one considers the first order variation of (\ref{problem}) with respect to $J$ at $J=0$ .    We have

\begin{eqnarray}
  \label{am1}
\frac{\delta}{\delta J} S[\phi_{J}] \Big|_{J=0} &=& \frac{ \delta
S_0}{\delta J}\Big|_{J=0} + \lim_{t_+ \rightarrow + \infty}
 \int_{\Sigma_{t_+}}  \sqrt{g_{\Sigma_{t_+}}}  (\phi_0 n_{t_+}^a
\partial_a K_0 + K_0 n_{t_+}^a \partial_a \phi) \cr
 &+&   \lim_{t_- \rightarrow - \infty}
 \int_{\Sigma_{t_-}}  \sqrt{g_{\Sigma_{t_-}}}  (\phi_0 n_{t_-}^a
\partial_a K_0 + K_0 n_{t_-}^a \partial_a \phi_0).
\end{eqnarray}

For example, we may consider the special case where the background solution $\phi_{0}$ has only one mode of the field excited: $\phi_{0} =  \cos(\omega t) f(x) $, for some fixed spatial profile $f(x)$.
As a Green's function, $K_0$ includes some non-zero contribution from each mode of $\phi$ in the AdS space and will typically excite each such mode to both the past and future of the region in which the boundary conditions are varied.  Thus, the integral over, say, $\Sigma_{t_+}$ will give a non-zero result of the form $A \cos^2(\omega t_+) + B \cos(\omega t_+)\sin(\omega t_+).$  This certainly does not converge as $t_+$ is taken to $+ \infty$.

Similar ambiguities arise even the trivial background $\phi_0=0$.  Consider, for example, the CFT two-point function.  Following the same steps that led to (\ref{am1}) and using $\phi_0=0$, one finds

\begin{eqnarray}
  \label{am2}
\frac{\delta}{\delta J(x)} \frac{\delta}{\delta J(y)} S[\phi_{J}] \Big|_{J=0} &=&
\frac{\delta}{\delta J(x)} \frac{\delta}{\delta J(y)} S_0\Big|_{J=0}  + 2 \lim_{t_+ \rightarrow + \infty}
 \int_{\Sigma_{t_+}}  \sqrt{g_{\Sigma_{t_+}}}  K_0 n_{t_+}^a
\partial_a K_0
\cr  &+&   2 \lim_{t_- \rightarrow - \infty}
 \int_{\Sigma_{t_-}}  \sqrt{g_{\Sigma_{t_-}}}  K_0 n_{t_-}^a
\partial_a K_0,
\end{eqnarray}
where the boundary terms at $t_\pm$ are quasi-periodic in $t_\pm$ for any choice of $K_0$ and cannot vanish identically since $K_0 \neq 0$.

\end{document}